\pdfoutput=1
\newcommand*{\papertemplate}{acm-report} % acm-report (polished), acm
\newif\ifpaperacm
\newif\ifpaperacmreport
\newif\ifpaperieee
\newif\ifpaperusenix
\def\acmtemplate{acm}
\def\acmreporttemplate{acm-report}
\def\ieeetemplate{ieee}
\def\usenixtemplate{usenix}
\ifx\papertemplate\acmtemplate
  \paperacmtrue
\else\ifx\papertemplate\acmreporttemplate
  \paperacmtrue
  \paperacmreporttrue
\else\ifx\papertemplate\ieeetemplate
  \paperieeetrue
\else\ifx\papertemplate\usenixtemplate
  \paperusenixtrue
\else
  \errmessage{Unknown paper template: use acm-report, acm, ieee, or usenix}
\fi\fi\fi\fi

\ifpaperacmreport
\documentclass[sigconf,nonacm,balance=false]{Template/acmart-primary/acmart}
\else\ifpaperacm
\documentclass[sigconf]{Template/acmart-primary/acmart}
\else\ifpaperieee
\documentclass[conference,letterpaper]{Template/IEEEtran}
\else
\documentclass[letterpaper,twocolumn,10pt]{article}
\fi\fi\fi
\ifpaperacm
  \ifpaperacmreport
    \setcopyright{none}
  \fi
  \newcommand{\paperbibliographystyle}{Template/acmart-primary/ACM-Reference-Format}
\else\ifpaperieee
  \usepackage{cite}
  \newcommand{\paperbibliographystyle}{IEEEtran}
\else
  \usepackage{Template/usenix2019_v3}
  \newcommand{\paperbibliographystyle}{plainurl}
\fi\fi

\usepackage{tikz}
\usepackage{amsmath}

\usepackage{enumitem}
\ifpaperieee\else
\usepackage{subcaption}
\fi
\ifpaperacm\else
\usepackage{amssymb,amsfonts}
\fi
\usepackage{algorithmic}
\usepackage{graphicx}
\usepackage{float}
\usepackage{textcomp}
\usepackage{xcolor}
\def\BibTeX{{\rm B\kern-.05em{\sc i\kern-.025em b}\kern-.08em
T\kern-.1667em\lower.7ex\hbox{E}\kern-.125emX}}

\makeatletter
\newcommand{\parheading}[1]{%
  \par
  \if@nobreak\else\addvspace{\baselineskip plus 2pt}\fi
  \noindent\textbf{#1}%
}
\makeatother
\newcommand{\todo}[1]{\textcolor{red}{\textbf{[TODO: #1]}}}

\newcommand{\append}[1]{#1}

\usepackage{array,multirow}
\newcommand{\band}[2]{\multirow{#1}{*}{\shortstack{#2}}}
\usepackage[most]{tcolorbox}
\usepackage{tikz}
\usetikzlibrary{positioning,arrows.meta,calc,fit,backgrounds}

\newtcolorbox{stickynote}[1][Note]{
  colback=yellow!8, colframe=black!60,
  boxrule=0.4pt, arc=1pt,
  left=4pt, right=4pt, top=3pt, bottom=3pt,
  title=#1, fonttitle=\bfseries\footnotesize,
  coltitle=black, colbacktitle=yellow!25,
  fontupper=\footnotesize\raggedright,
  before skip=4pt, after skip=4pt
}

\definecolor{cInfo}{HTML}{2563EB}  % blue   — general information
\definecolor{cEdge}{HTML}{D97706}  % amber  — edge cases
\definecolor{cWarn}{HTML}{DC2626}  % red    — disclaimers/pitfalls
\definecolor{cTip}{HTML}{059669}   % green  — tips/takeaways

\tcbset{accentnote/.style 2 args={
    enhanced, frame hidden,
    colback=#2!6!white,
    borderline west={2.5pt}{0pt}{#2},
    arc=1pt, left=6pt, right=6pt, top=4pt, bottom=4pt,
    before skip=6pt, after skip=6pt,
    fontupper=\footnotesize,
    fonttitle=\bfseries\footnotesize, coltitle=#2!70!black,
    detach title, before upper={\tcbtitle\enspace},
    title={#1.}
}}

\newtcolorbox{info}[1][Info]{accentnote={#1}{cInfo}}
\newtcolorbox{edgecase}[1][Edge case]{accentnote={#1}{cEdge}}
\newtcolorbox{disclaimer}[1][Disclaimer]{accentnote={#1}{cWarn}}
\newtcolorbox{tip}[1][Tip]{accentnote={#1}{cTip}}
\usepackage{listings}

\definecolor{codegreen}{rgb}{0,0.6,0}
\definecolor{codegray}{rgb}{0.5,0.5,0.5}
\definecolor{codepurple}{rgb}{0.58,0,0.82}
\definecolor{backcolour}{rgb}{0.95,0.95,0.92}
\definecolor{commentgray}{gray}{0.45}

\lstdefinestyle{cppstyle}{
  language=C++,
  backgroundcolor=\color{backcolour},
  commentstyle=\color{commentgray},
  keywordstyle=\color{magenta},
  numberstyle=\tiny\color{codegray},
  stringstyle=\color{codepurple},
  basicstyle=\ttfamily\footnotesize,
  breakatwhitespace=false,
  breaklines=true,
  captionpos=b,
  keepspaces=true,
  numbers=left,
  numbersep=5pt,
  showspaces=false,
  showstringspaces=false,
  showtabs=false,
  tabsize=2,
}
\usepackage{booktabs}
\usepackage{array,tabularx}
\newcolumntype{Y}{>{\raggedright\arraybackslash}X}
\usepackage{siunitx}
\ifpaperacm
\newcommand{\tablefigures}{\fontfamily{LinuxLibertineT-TLF}\selectfont}
\else
\newcommand{\tablefigures}{}
\fi
\AtBeginEnvironment{tabular}{\tablefigures}
\AtBeginEnvironment{tabularx}{\tablefigures}
\AtBeginEnvironment{table}{\footnotesize\setlength{\tabcolsep}{3pt}}
\AtBeginEnvironment{table*}{\footnotesize\setlength{\tabcolsep}{3pt}}
\newcommand{\tabgroup}[2]{\multicolumn{#1}{@{}l}{\emph{\textbf{#2}}}\\}
\usepackage{comment}
\usepackage{xurl}

\ExplSyntaxOn
\NewDocumentCommand{\code}{m}
{
  \texttt
  {
    \tl_set:Nn \l_tmpa_tl { #1 }
    \tl_replace_all:Nen \l_tmpa_tl { \char_generate:nn { `_ } { 8 } } { \_ }
    \tl_use:N \l_tmpa_tl
  }
}
\ExplSyntaxOff
\ifpaperacmreport
  \newlength{\paperauthorgap}
  \makeatletter
  \AddToHook{cmd/@mkauthors/after}{%
    \global\setbox\mktitle@bx=\vbox{%
      \unvbox\mktitle@bx\vskip\paperauthorgap}%
  }
  \makeatother

  \makeatletter\@beginparpenalty=10000\makeatother
  \AtBeginEnvironment{table}{\renewcommand{\arraystretch}{1.12}}
  \AtBeginEnvironment{table*}{\renewcommand{\arraystretch}{1.12}}

  \definecolor{paperlink}{HTML}{285F78}
  \AtEndPreamble{\hypersetup{colorlinks=true,linkcolor=black,
    citecolor=black,urlcolor=paperlink}}

  \makeatletter
  \renewcommand{\parheading}[1]{%
    \par
    \if@nobreak\else\addvspace{0.55\baselineskip}\fi
    \noindent\textbf{#1}%
  }
  \makeatother
\fi

\newcommand{\PaperTitle}{GPU-Initiated Communication: Dissecting Down
to the Bone}
\title{\PaperTitle}

\newcommand{\GitHubRepo}{https://github.com/ParCoreLab/Dissecting-GPU-Communication-Experiments}
\newcommand{\GitHubRepoURL}{\url{\GitHubRepo}}

\newcommand{\paperkeywords}{GPU-initiated communication, GPU-initiated
  networking, RDMA, IBGDA, GDAKI, GPUDirect Async, CPU proxies, NVSHMEM,
  NCCL GIN, DeepEP, Mixture-of-Experts, expert parallelism, InfiniBand,
ConnectX-7, performance characterization}
\ExplSyntaxOn
\seq_new:N \g_paper_authors_seq

\cs_new_protected:Npn \paper_author_details:nnnnn #1#2#3#4#5
  {
    \tl_if_blank:nF {#1} {#1\\}
    #2
    \tl_if_blank:nF {#3#4}
      {
        \\
        #3
        \tl_if_blank:nF {#3} {\tl_if_blank:nF {#4} {,~}}
        #4
      }
    \tl_if_blank:nF {#5} {\\{\normalfont\texttt{#5}}}
  }

\NewDocumentCommand{\PaperAuthor}{O{} m m m m m m}
  {
    \ifpaperacm
      \author{#2}
      \tl_if_blank:nF {#1} {\orcid{#1}}
      \affiliation{
        \tl_if_blank:nF {#3} {\department{#3}}
        \institution{#4}
        \city{#5}
        \country{#6}
      }
      \tl_if_blank:nF {#7} {\email{#7}}
    \else
      \seq_gput_right:Nn \g_paper_authors_seq
        {
          \ifpaperieee
            \IEEEauthorblockN{#2}
            \IEEEauthorblockA{
              \paper_author_details:nnnnn {#3}{#4}{#5}{#6}{#7}
            }
          \else
            {\normalfont #2}\\
            \paper_author_details:nnnnn {#3}{#4}{#5}{#6}{#7}
          \fi
        }
      \author{\seq_use:Nn \g_paper_authors_seq {\and}}
    \fi
  }
\ExplSyntaxOff

\PaperAuthor{Javid Baydamirli}
{} % Department name
{Ko\c{c} University, Türkiye}
{}
{}
{jbaydamirli21@ku.edu.tr}

\PaperAuthor{Ismayil Ismayilov}
{}
{fal}
{}
{}
{ismayil@fal.ai}

\PaperAuthor{Kaan Oktay}
{}
{fal}
{}
{}
{kaan@fal.ai}

\PaperAuthor{Didem Unat}
{}
{Ko\c{c} University, Türkiye}
{}
{}
{dunat@ku.edu.tr}

\date{}

\begin{document}

% ACM collects the abstract before making the full-width title block.
\ifpaperacm
\begin{abstract}
  GPU-initiated communication lets GPU threads post RDMA operations
  directly to the NIC. It underpins NVSHMEM, NCCL GIN, and DeepEP,
  which serve the fine-grained, latency-critical communication of
  Mixture-of-Experts (MoE) models, yet its performance characteristics
  and optimizations remain scarcely documented beyond source code, and
  library comparisons fail to separate the costs of the hardware
  mechanism from those of the library around it.

  This paper dissects GPU-initiated communication at the GPU--NIC
  boundary. We first detail the GPU-side network path: queue placement,
  work-request construction, doorbell ordering, and completion
  semantics. We then introduce mini-gda and mini-proxy, minimal
  transports for the GPU and CPU-proxy submission paths, and measure
  them alongside NVSHMEM IBGDA, NCCL GIN, DeepEP, UCCL-EP, MSCCL++, and
  fabric-lib on NVIDIA H100, H200, B200, and GB200 platforms.
  A minimal GPU path issues an operation in \SI{0.7}{\micro\second} and
  completes in \SI{4.0}{\micro\second}; libraries add up to
  \SI{4.6}{\micro\second} of issue time through queue management,
  memory ordering, and completion scope, and issue time scales with the
  SM clock. A tuned CPU proxy matches or beats the GPU path at idle, at
  the cost of a dedicated core whose operating state sets its latency
  and throughput. On either path, sharing a queue with bulk traffic
  raises latency by one to three orders of magnitude. Reaching the
  260\,M\,msg/s ceiling of our InfiniBand platform requires doorbell
  batching and queue parallelism, and both have resource costs:
  communication code can reduce GPU block residency even when unused,
  and all-to-all traffic loses 59\% of its NIC message rate at about
  3,000 active connections. The submission path alone therefore does
  not predict communication performance. Our experiment code and
  results are available at \GitHubRepoURL.
\end{abstract}

\maketitle
\fi
% PDF metadata for indexers (Google Scholar reads it). Set
% after \maketitle because acmart's \maketitle overwrites these fields.
% hyperref drops the author and keyword fields in this setup (they never
% reach the PDF, even as package options), so those two go through
% pdfTeX's \pdfinfo directly. Guarded because not every template loads
% hyperref and not every engine has \pdfinfo.
\ifdefined\hypersetup
\hypersetup{
  pdftitle={\PaperTitle},
  pdfsubject={Performance characterization of GPU-initiated RDMA
  communication and CPU proxies}
}
\fi
\ifdefined\pdfinfo
\pdfinfo{
  /Author (Javid Baydamirli, Ismayil Ismayilov, Kaan Oktay, Didem Unat)
  /Keywords (GPU-initiated communication, GPU-initiated networking, RDMA,
    IBGDA, GDAKI, GPUDirect Async, CPU proxies, NVSHMEM, NCCL GIN, DeepEP,
    Mixture-of-Experts, expert parallelism, InfiniBand, ConnectX-7,
  performance characterization)
}
\fi
\ifpaperacm
\else
\maketitle
\begin{abstract}
  GPU-initiated communication lets GPU threads post RDMA operations
  directly to the NIC. It underpins NVSHMEM, NCCL GIN, and DeepEP,
  which serve the fine-grained, latency-critical communication of
  Mixture-of-Experts (MoE) models, yet its performance characteristics
  and optimizations remain scarcely documented beyond source code, and
  library comparisons fail to separate the costs of the hardware
  mechanism from those of the library around it.

  This paper dissects GPU-initiated communication at the GPU--NIC
  boundary. We first detail the GPU-side network path: queue placement,
  work-request construction, doorbell ordering, and completion
  semantics. We then introduce mini-gda and mini-proxy, minimal
  transports for the GPU and CPU-proxy submission paths, and measure
  them alongside NVSHMEM IBGDA, NCCL GIN, DeepEP, UCCL-EP, MSCCL++, and
  fabric-lib on NVIDIA H100, H200, B200, and GB200 platforms.
  A minimal GPU path issues an operation in \SI{0.7}{\micro\second} and
  completes in \SI{4.0}{\micro\second}; libraries add up to
  \SI{4.6}{\micro\second} of issue time through queue management,
  memory ordering, and completion scope, and issue time scales with the
  SM clock. A tuned CPU proxy matches or beats the GPU path at idle, at
  the cost of a dedicated core whose operating state sets its latency
  and throughput. On either path, sharing a queue with bulk traffic
  raises latency by one to three orders of magnitude. Reaching the
  260\,M\,msg/s ceiling of our InfiniBand platform requires doorbell
  batching and queue parallelism, and both have resource costs:
  communication code can reduce GPU block residency even when unused,
  and all-to-all traffic loses 59\% of its NIC message rate at about
  3,000 active connections. The submission path alone therefore does
  not predict communication performance. Our experiment code and
  results are available at \GitHubRepoURL.
\end{abstract}

\ifpaperieee
\else
\par
\fi
\fi

% Shown only with the USENIX template (NSDI).
\ifpaperusenix
\todo{NSDI camera-ready: the plots in \code{Figures/Experiments/} are set
  in Linux Libertine, the arXiv body font, but this template sets the body
  in Times. Set \code{FONT = "times"} in the experiments repo's
  \code{plots/figstyle.py}, rerun the \code{plots/paper} scripts, and copy
  the PDFs here. The TikZ figures (Figures 1 and 2) follow the body font
  on their own. Also check the 12-page limit: in this layout the body ran
to page 14 on 2026-09-30.}
\fi

% ================================================================
% Sections
\section{Introduction}
\label{sec:Introduction}

Communication across GPUs has historically been managed by the CPU, where a
host thread builds RDMA work requests, rings the NIC doorbell, and polls
for completion.
Modern GPU-initiated transports like IBGDA \cite{NVIDIA-Blog-IBGDA}
and GDAKI \cite{gpuinitiatednetworkingnccl} instead move this control
to the GPU, enabling GPU threads to construct RDMA descriptors
and ring NIC doorbells themselves, over PCIe.
Workloads with irregular, latency-critical communication
patterns benefit the most from this paradigm. In Mixture-of-Experts~(MoE)
models, GPU kernels decide which experts each token is sent to, so
transfers are fine-grained and data-dependent, and network delay stalls
the GPU directly~\cite{deepep2025,pplx-blog}.
This pattern has motivated several specialized communication libraries
like DeepEP \cite{deepep2025} and pplx-kernels \cite{pplx-kernels}
that build on top of GPU-initiated NVSHMEM and NCCL GIN to reduce
the latency of dispatch and combine kernels
\cite{Nvshmem_Docs, pplx-blog}.
Conversely, the same workload is also targeted by libraries that
opt to keep the host on the communication path. UCCL-EP \cite{uccl-ep},
pplx-garden/fabric-lib \cite{pplx-garden-fabric-lib}, and
MSCCL++ PortChannel \cite{msccl++} forgo GPU submission
to maintain cross-vendor GPU and NIC portability, while still
reporting low latency with workload-specific proxy designs.\looseness=1

\begin{table*}[t]
  \centering
  \caption{Queue, posting, and completion choices of the implementations
    we measure. Worker and queue counts are benchmark settings.
    PE: processing element (one GPU rank); DCI: dynamically connected
    initiator (\S\ref{sec:boundary:transport}); LL: DeepEP's
    low-latency kernels.
    Versions: NVSHMEM 3.4.5 and 3.7.2, NCCL 2.30.7 and 2.31.2
    (Table~\ref{tab:platforms}), UCCL-EP
  \code{a3d520e}, MSCCL++ \code{6231b4f}, fabric-lib \code{2446003}.}
  \label{tab:impls}
  \begin{tabularx}{\textwidth}{@{}>{\raggedright\arraybackslash}p{0.14\textwidth}
      >{\raggedright\arraybackslash}p{0.13\textwidth} Y
      >{\raggedright\arraybackslash}p{0.18\textwidth}
    >{\raggedright\arraybackslash}p{0.19\textwidth}@{}}
    \toprule
    \textbf{Implementation} & \textbf{Submitter} & \textbf{Queues and
    workers} & \textbf{Posting batch} &
    \textbf{Completion scope} \\
    \midrule
    NVSHMEM IBGDA & GPU & RC per peer or a shared DCI pool
    & default 32; earlier if alone &
    all configured QPs \\
    DeepEP V1 & GPU & LL kernels assign RC QPs to local experts
    & every 4th message per expert & one $(\text{peer},\text{QP})$ \\
    NCCL GIN GDAKI & GPU & explicit contexts with per-peer RC connections
    & caller-controlled aggregation & local or context flush \\
    mini-gda & GPU & 1--528 QPs, 1--16,384 threads
    & 1--256 WQEs per doorbell & selected QP progress \\
    \midrule
    NVSHMEM IBRC & one CPU proxy & one descriptor FIFO per PE
    & library-managed & proxy quiet \\
    NCCL GIN Proxy & 1--4 CPU workers & per-context descriptor rings
    & library-managed & context flush \\
    UCCL-EP & 1--8 CPU workers & 8 FIFOs per worker; MSCCL++ FIFO backend
    & adaptive chains & channel local \\
    MSCCL++ & 1--8 CPU services & one FIFO per service, 16\,B entries
    & one request & service local \\
    fabric-lib & one CPU/NIC & host-initiated native
    client in this study
    & up to four & transfer local \\
    mini-proxy & 1--8 CPU workers & 1--32 host rings, 16\,B entries
    & 1--16 & host or GPU counter \\
    \bottomrule
  \end{tabularx}
\end{table*}

Published measurements favor each design on a different metric.
Proxies report lower latency for a single unloaded operation, and
GPU paths higher rates for concurrent small messages
\cite{gpuinitiatednetworkingnccl,pplx-garden-fabric-lib,NVIDIA-Blog-IBGDA}.
These studies
\cite{NVIDIA-Blog-IBGDA,gpuinitiatednetworkingnccl,deepep2025,pplx-blog,uccl-ep,msccl++},
however, typically
evaluate libraries as a whole, conflating higher-level software
wrappers with the underlying hardware mechanisms.
A library-level comparison obscures whether the performance
differences come from queue counts, batching strategies, memory
fence strengths, or other API overheads, even when said libraries
drive the same hardware interface. It may compare an implementation
with a single queue against one with sixteen, or a completion over
all available queues against one for a single peer.\looseness=1

We therefore isolate each mechanism from the libraries and ask three
questions:
\begin{enumerate}
  \item \textbf{What is the cost of a single operation?}
    What does one GPU-initiated RDMA write cost, and how do
    WQE construction, queue management, memory ordering,
    completion scope, and the processor clock contribute?
  \item \textbf{When does GPU or proxy submission perform better?}
    How do proxy design choices such as workers, batching, and queue
    sharing dictate latency and throughput, and how do the two paths
    compare under different load conditions?
  \item \textbf{What sustains high message rates, and what does it cost?}
    How do GPU threads, NIC queues, doorbell batching, and warp and SM
    placement contribute to throughput, and what do they cost in GPU
    block residency and NIC connection state?
\end{enumerate}

To answer these questions, we implement two minimal transport drivers:
\textbf{mini-gda} for the GPU-submission path, and \textbf{mini-proxy}
for the CPU-proxy path. They perform only the steps the hardware
requires and parametrize the design choices,
which lets us
evaluate each mechanism in isolation without an attached library.
Alongside them,
we evaluate several production libraries (Table~\ref{tab:impls}) on
NVIDIA H100, H200, B200, and GB200 platforms with ConnectX-7 NICs
over InfiniBand and RoCEv2.\looseness=1

Our contributions are the following:
\begin{itemize}
  \item A mechanism-level description of the GPU--NIC boundary
    (\S\ref{sec:GPU-NIC-Boundary}), covering where GPU-resident RDMA
    queues live, how device code constructs work requests, and the
    doorbell, ordering, and completion semantics that libraries
    implement differently.
  \item \textbf{mini-gda} and \textbf{mini-proxy}, two minimal
    transports that isolate the mechanism from the library, together
    with a microbenchmark suite built on them
    (\S\ref{sec:Microbenchmarks}). We release both as
    \href{\GitHubRepo}{open source}.
  \item An evaluation across NVIDIA H100, H200, B200, and GB200
    platforms (\S\ref{sec:Microbenchmarks}) that quantifies
    single-operation cost, proxy trade-offs, message rate, and the
    resource costs of the transport, and identifies the configuration
    choices behind divergent performance across libraries.
\end{itemize}

The evaluation yields three main findings.
\begin{enumerate}
  \item \textbf{Software overhead and completion semantics determine
    single-operation latency.}
    A minimal GPU path issues an 8\,B write in \SI{0.7}{\micro\second}
    and completes in \SI{4.0}{\micro\second}. Communication libraries
    add varying costs on top through their queue handling, memory
    ordering, and completion scope: GDAKI takes
    \SI{1.6}{\micro\second} to issue and NVSHMEM's public API
    \SI{5.3}{\micro\second}. Ordering scope alone changes issue time
    $3.7\times$ across safe configurations, and completing all
    configured queues doubles put+completion latency from 1 to 16
    queues. Issue latency scales strongly with the SM clock.
  \item \textbf{GPU path vs.\ CPU proxy depends on latency, capacity,
    and isolation needs.}
    A tuned proxy completes within \SI{0.07}{\micro\second} of the
    minimal GPU path and has a lower median round trip (5.9
    vs.\ \SI{6.9}{\micro\second}), at the cost of a dedicated core whose
    operating state determines its results. Shared queues degrade under
    load for both submission paths: a proxy FIFO shared with bulk
    traffic reaches tens of milliseconds, and a reserved queue recovers
    one to two orders of magnitude. More workers and batching raise the
    proxy message rate, which stays well below IBGDA's on our
    InfiniBand platform but reaches 90\% of it on GB200. Every path
    except a cold IBRC proxy exceeds 90\% of line rate by 7\,KiB, the
    size of a DeepSeek-V3 token's FP8 hidden state.
  \item \textbf{High message rates require queue scaling and doorbell
    batching, but both cost resources.}
    One submitting thread posts 1.8\,M\,msg/s. Cooperative publication
    lets a warp reach 21.6\,M\,msg/s on one queue, and independent
    queues raise GPU submission throughput to 260\,M\,msg/s. GPU
    submission is not required for this peak: a host thread ringing
    doorbells for GPU-built WQEs reaches 257--258\,M\,msg/s.
    Communication code can cost up to 37\% of a kernel's useful
    throughput even when dormant, by lowering block residency. At the
    NIC, active connections cost more when sending and receiving
    together: a send-only NIC keeps about 242\,M\,msg/s through 32,768
    active queues, while all-to-all traffic loses 59\% of its rate near
    3,000 connections, and Dynamically Connected (DC) transport does not
    remove this decline.
\end{enumerate}

Section~\ref{sec:Background} introduces GPU-initiated communication
and the libraries we measure, Section~\ref{sec:GPU-NIC-Boundary}
describes the GPU--NIC boundary, Section~\ref{sec:Microbenchmarks}
presents the evaluation, Section~\ref{sec:Related-Work} discusses
related work, and Section~\ref{sec:Conclusion} concludes with
guidelines for library designers and benchmarkers.

\section{Background}
\label{sec:Background}

\parheading{GPUDirect.} GPUDirect is a family of technologies that
progressively removes the host
from the networking path. GPUDirect RDMA \cite{gpudirect-rdma}
exposes GPU memory over PCIe for direct NIC access, removing
host-side staging copies.
GPUDirect Async \cite{GPUDirect-Async} shifts synchronization onto
the device, letting GPUs trigger and wait on communication
pre-posted by the CPU. Its kernel-initiated successors, IBGDA
\cite{Nvshmem_Docs-2.6.0} and GDAKI
\cite{gpuinitiatednetworkingnccl}, move submission entirely into the
kernel, enabling GPU threads to build queue elements, ring doorbells
over PCIe, and poll completions natively.

Two auxiliary components are important to note: GDRCopy \cite{gdrcopy}, which
gives the CPU low-latency load/store access to GPU memory and underpins
CPU-assisted proxy paths, and \code{rdma-core}'s \code{mlx5dv} direct
verbs \cite{rdma-core}, with which user space creates the queues,
doorbell pages, and memory registrations that transport libraries place
in GPU memory.

\parheading{RDMA Communication.}
RDMA communication is organized around Queue Pairs (QPs), each consisting of a
Send Queue (SQ) and a Receive Queue (RQ), both structured as circular rings of
Work Queue Elements (WQEs).
Send and receive completions are directed
to completion queues (CQs), which can be distinct or shared with other
QPs.
The NIC reports a finished \emph{signaled} WQE by writing a completion
queue entry (CQE), while unsignaled WQEs produce no CQE of their own.

\parheading{Communication Control vs.\ Submission.}
We distinguish the agent that \emph{semantically decides} what to
communicate from the agent that \emph{physically submits} work to the
network \cite{Landscape_Of_GPU_Comm}.
The operation is initiated either by
a CPU call (host-initiated), as in classic MPI/NCCL collectives,
or by device code (GPU-initiated). A GPU-initiated operation may
then be \emph{GPU-submitted}, where the GPU writes the doorbell itself,
or \emph{proxy-submitted}, where the request is forwarded to
a CPU proxy thread.
Additionally, WQEs can be constructed either by the
CPU or the GPU, irrespective of the submitter: some implementations
delegate only the doorbell submission to the CPU to bypass platform
limitations (\S\ref{sec:boundary:geography}).

Table~\ref{tab:impls} summarizes
the queue, posting, and completion choices of the GPU-submitted and
proxy-submitted libraries we use in our evaluation.

\section{The GPU--NIC Boundary}
\label{sec:GPU-NIC-Boundary}
\begin{figure}[t]
  \centering
  % GPU-submitted RDMA write as a sequence diagram (fig:gpu-rdma-path).
% The doorbell record has its own lifeline in GPU memory, and the
% ordering rule sits next to the fence. \input{Figures/gpu_nic_dbrec} at
% natural size (column width). Uses arrows.meta, loaded in main.tex.
\input{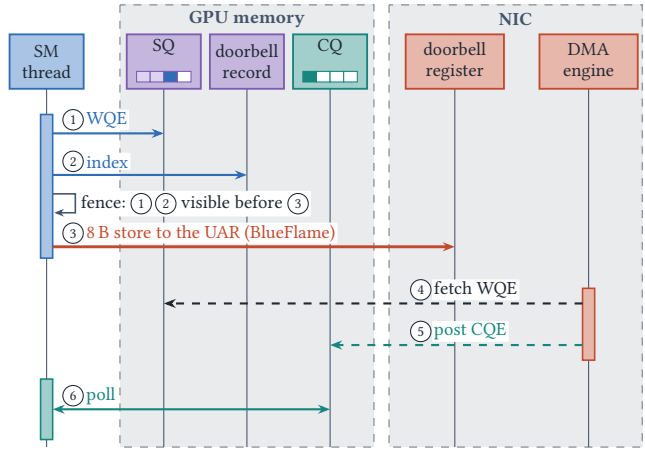}%
\begin{tikzpicture}[x=1cm, y=1cm, line width=0.5pt,
  font=\fontsize{7}{8}\selectfont,
  head/.style={draw=#1, line width=0.7pt, minimum width=0.98cm, minimum height=0.72cm,
    align=center, inner sep=1pt, text=figink},
  life/.style={figink2, line width=0.4pt},
  panel/.style={fill=figpanel},
  pframe/.style={draw=figmuted, dashed, line width=0.5pt},
  ptitle/.style={font=\fontsize{7}{8}\selectfont\bfseries, text=figink2},
  num/.style={circle, anchor=center, draw=figink, line width=0.4pt, fill=white, inner sep=0pt,
    minimum size=2.7mm, font=\fontsize{5.5}{6}\selectfont, text=figink},
  msg/.style={-{Stealth[length=1.6mm, width=1.2mm]}, line width=0.8pt},
  lbl/.style={fill=white, inner sep=0.8pt, rounded corners=0.5pt},
  act/.style={draw=#1, line width=0.6pt}]
  \def\xsm{0.55}\def\xsq{2.1}\def\xdr{3.2}\def\xcq{4.3}\def\xdb{5.95}\def\xdma{7.72}
  \fill[panel] (1.52,0.3) rectangle (4.88,-5.55);
  \fill[panel] (5.08,0.3) rectangle (8.43,-5.55);
  \draw[pframe] (1.52,0.3) rectangle (4.88,-5.55);
  \draw[pframe] (5.08,0.3) rectangle (8.43,-5.55);
  \node[ptitle] at (3.2,0.12) {GPU memory};
  \node[ptitle] at (6.755,0.12) {NIC};
  \foreach \x in {\xsm,\xsq,\xdr,\xcq,\xdb,\xdma} \draw[life] (\x,-0.55) -- (\x,-5.55);
  \node[head=figblue, fill=figmblue] at (\xsm,-0.45) {SM\\thread};
  \node[head=figpurp, fill=figmpurp] at (\xsq,-0.45) {SQ\\[1.6mm]};
  \node[head=figpurp, fill=figmpurp] at (\xdr,-0.45) {doorbell\\record};
  \node[head=figteal, fill=figmteal] at (\xcq,-0.45) {CQ\\[1.6mm]};
  \node[head=figterra, fill=figmterra, minimum width=1.3cm] at (\xdb,-0.45) {doorbell\\register};
  \node[head=figterra, fill=figmterra, minimum width=1.3cm] at (\xdma,-0.45) {DMA\\engine};
  \foreach \i/\c in {0/figtpurp,1/figtpurp,2/figblue,3/white}
    \filldraw[draw=figpurp, fill=\c, line width=0.4pt] (\xsq-0.36+\i*0.18,-0.72) rectangle ++(0.18,0.14);
  \foreach \i/\c in {0/figteal,1/white,2/white,3/white}
    \filldraw[draw=figteal, fill=\c, line width=0.4pt] (\xcq-0.36+\i*0.18,-0.72) rectangle ++(0.18,0.14);
  % activation bars
  \filldraw[act=figblue, fill=figmblue] (\xsm-0.08,-1.15) rectangle (\xsm+0.08,-3.05);
  \filldraw[act=figterra, fill=figmterra] (\xdma-0.08,-3.45) rectangle (\xdma+0.08,-4.4);
  \filldraw[act=figteal, fill=figmteal] (\xsm-0.08,-4.65) rectangle (\xsm+0.08,-5.45);
  % (1) WQE into the SQ, (2) doorbell record
  \draw[msg, figblue] (\xsm+0.08,-1.4) -- (\xsq,-1.4);
  \node[num] (n1) at (0.9,-1.22) {1}; \node[lbl, anchor=west, text=figblue] at (n1.east) {WQE};
  \draw[msg, figblue] (\xsm+0.08,-1.95) -- (\xdr,-1.95);
  \node[num] (n2) at (0.9,-1.77) {2}; \node[lbl, anchor=west, text=figblue] at (n2.east) {index};
  % fence: (1) and (2) visible before (3)
  \draw[-{Stealth[length=1.3mm, width=1mm]}, figink2, line width=0.6pt]
    (\xsm+0.08,-2.2) -- ++(0.3,0) -- ++(0,-0.25) -- (\xsm+0.1,-2.45);
  \node[lbl, anchor=west, text=figink] at (\xsm+0.42,-2.33)
    {fence: \tikz[baseline=-0.6ex]\node[num]{1};\,\tikz[baseline=-0.6ex]\node[num]{2}; visible before \tikz[baseline=-0.6ex]\node[num]{3};};
  % (3) doorbell store
  \draw[msg, figterra, line width=1.1pt] (\xsm+0.08,-2.9) -- (\xdb,-2.9);
  \node[num] (n3) at (0.9,-2.72) {3}; \node[lbl, anchor=west, text=figterra] at (n3.east) {8\,B store to the UAR (BlueFlame)};
  % (4) fetch WQE, (5) post CQE
  \draw[msg, figink, dashed] (\xdma-0.08,-3.65) -- (\xsq,-3.65);
  \node[num] (n4) at (5.5,-3.47) {4}; \node[lbl, anchor=west, text=figink] at (n4.east) {fetch WQE};
  \draw[msg, figteal, dashed] (\xdma-0.08,-4.2) -- (\xcq,-4.2);
  \node[num] (n5) at (5.5,-4.02) {5}; \node[lbl, anchor=west, text=figteal] at (n5.east) {post CQE};
  % (6) poll
  \draw[{Stealth[length=1.6mm, width=1.2mm]}-{Stealth[length=1.6mm, width=1.2mm]}, figteal, line width=0.8pt]
    (\xsm+0.08,-5.05) -- (\xcq,-5.05);
  \node[num] (n6) at (0.9,-4.87) {6}; \node[lbl, anchor=west, text=figteal] at (n6.east) {poll};
\end{tikzpicture}%
  \caption{The GPU-submitted path with GPU-resident queues and
    in-kernel WQE construction. In a CPU-proxy path the host performs
    (1)--(3) and (6), and the
  GPU instead enqueues a request descriptor.}
  \label{fig:gpu-rdma-path}
\end{figure}

A network transfer requires NIC queues the submitter can reach, WQE
construction, ordered doorbell updates, and completion polling
(Figure~\ref{fig:gpu-rdma-path}), which we explain in
turn.
Libraries implement each step differently, and we measure
the respective costs of each choice in Section~\ref{sec:Microbenchmarks}.

\subsection{Memory geography}
\label{sec:boundary:geography}
In traditional CPU-side RDMA, the SQ and RQ WQE buffers, the completion
queue, and the doorbell record typically live in host memory. As GPU submission
requires device access to these RDMA queues, implementations like IBGDA and
GDAKI allocate them in GPU-accessible memory (typically with \code{cudaMalloc}),
where SMs populate and modify them with loads and stores.
The NIC reaches the buffers through GPUDirect RDMA mappings
established at registration time with \code{nvidia-peermem} or DMA-BUF
\cite{gpudirect-rdma-ordering,nvshmem-host-source}.

While the queues can live in any memory subsystem the NIC can reach by DMA,
the doorbell register cannot be freely moved. It is a hardware register
inside a \emph{User Access Region} (UAR), which is a slice of the NIC's PCIe
BAR through which
user-space processes submit doorbells. The UAR must be mapped into the GPU's
virtual address space for kernels to submit doorbells directly. This is achieved
by allocating the UAR with \code{mlx5dv\_devx\_alloc\_uar},
and registering its BAR page with the CUDA driver as I/O memory
(\code{cuMemHostRegister}), which can then be used as an opaque
device pointer that SMs can store to (\code{cuMemHostGetDevicePointer})
\cite{nvshmem-host-source}.

% \begin{stickynote}[In practice]
%   In CUDA and \code{mlx5dv}, the transport allocates a UAR with
%   \code{mlx5dv\_devx\_alloc\_uar}, registers the resulting BAR page
% with the CUDA
%   driver through \code{cuMemHostRegister} with
%   \code{CU\_MEMHOSTREGISTER\_IOMEMORY}, and obtains a device pointer with
%   \code{cuMemHostGetDevicePointer} that SMs can store to.
% \end{stickynote}

% \begin{stickynote}[Mapping the doorbell]
Mapping a third-party PCIe BAR into GPU address space is a potential
security hazard, and the NVIDIA kernel module permits it only with the
option \code{PeerMappingOverride=1} \cite{nvshmem-30-blog}.
DMA-BUF registration lets the NIC access GPU memory but does not by itself
establish this reverse mapping \cite{nvshmem-host-source}. NVSHMEM's
\emph{CPU-assisted} mode keeps GPU-built WQEs and has a host thread
forward the doorbell value to the UAR for systems where the mapping is
disallowed \cite{nvshmem-30-blog}.
% This is selected with
% \code{NVSHMEM_IBGDA_NIC_HANDLER}: \code{cpu} uses GDRCopy to read
% GPU-resident progress, while \code{cpu_host_memory} uses shared host
% memory. Version 3.4.5 added the latter fallback without a GDRCopy
% dependency \cite{nvshmem-345-notes,nvshmem-host-source}.
% Former sticky note "Mapping the doorbell" (same content) folded into
% the paragraph above, 2026-09-14.
% \end{stickynote}

\subsection{Work requests, submission, and doorbell semantics}
\label{sec:boundary:wqe}
\label{sec:boundary:doorbell}

\begin{figure}[t]
  \centering
  % mlx5 send WQEs for an RDMA write, drawn as whole WQEs on the 64 B
% basic-block grid, with the control segment expanded below.
% Used by fig:wqe-segments. \input{Figures/segment_blocks} at natural
% size (fits \columnwidth). Needs only tikz and xcolor. The caption spells
% out ctrl, AV, raddr and data.
\input{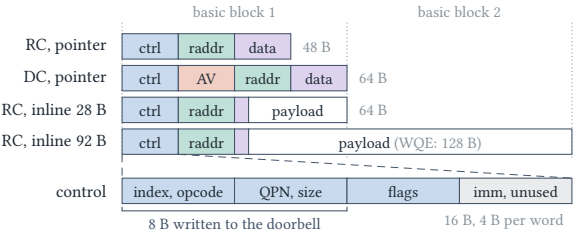}%
\begin{tikzpicture}[x=0.0465cm, y=0.42cm, line width=0.4pt,
  s/.style={draw=figink2},
  f/.style={font=\fontsize{6}{7}\selectfont, text=figink, text height=1.5ex, text depth=0.3ex},
  n/.style={font=\fontsize{6}{7}\selectfont, text=figmuted},
  lab/.style={font=\fontsize{6.5}{7.5}\selectfont, text=figink, anchor=east}]
  % basic-block grid
  \foreach \x in {0,64,128} \draw[figmuted, densely dotted] (\x,0.35) -- (\x,-3.9);
  \node[n, above] at (32,0.25) {basic block 1};
  \node[n, above] at (96,0.25) {basic block 2};
  \newcommand{\seg}[5]{\filldraw[fill=#3, s] (#1,#5) rectangle (#2,#5-0.8); \node[f] at ({(#1+#2)/2},#5-0.4) {#4};}
  \node[lab] at (-2,-0.4) {RC, pointer};
  \seg{0}{16}{figtblue}{ctrl}{0} \seg{16}{32}{figtteal}{raddr}{0} \seg{32}{48}{figtpurp}{data}{0}
  \node[n, anchor=west] at (49,-0.4) {48 B};
  \node[lab] at (-2,-1.4) {DC, pointer};
  \seg{0}{16}{figtblue}{ctrl}{-1} \seg{16}{32}{figtterra}{AV}{-1} \seg{32}{48}{figtteal}{raddr}{-1} \seg{48}{64}{figtpurp}{data}{-1}
  \node[n, anchor=west] at (65,-1.4) {64 B};
  \node[lab] at (-2,-2.4) {RC, inline 28 B};
  \seg{0}{16}{figtblue}{ctrl}{-2} \seg{16}{32}{figtteal}{raddr}{-2} \seg{32}{36}{figtpurp}{}{-2} \seg{36}{64}{white}{payload}{-2}
  \node[n, anchor=west] at (65,-2.4) {64 B};
  \node[lab] at (-2,-3.4) {RC, inline 92 B};
  \seg{0}{16}{figtblue}{ctrl}{-3} \seg{16}{32}{figtteal}{raddr}{-3} \seg{32}{36}{figtpurp}{}{-3} \seg{36}{128}{white}{payload \textcolor{figmuted}{(WQE: 128 B)}}{-3}
  % control segment, expanded
  \node[lab] at (-2,-4.95) {control};
  \foreach \a/\b/\t/\c in {0/32/{index, opcode}/figtblue, 32/64/{QPN, size}/figtblue, 64/96/flags/figtblue, 96/128/{imm, unused}/figpanel}
    {\filldraw[fill=\c, s] (\a,-4.55) rectangle (\b,-5.35); \node[f] at ({(\a+\b)/2},-4.95) {\t};}
  \draw[figink2, densely dashed, line width=0.3pt] (0,-3.8) -- (0,-4.55);
  \draw[figink2, densely dashed, line width=0.3pt] (16,-3.8) -- (128,-4.55);
  \draw[figink2, line width=0.4pt] (0,-5.45) -- (0,-5.58) -- (64,-5.58) -- (64,-5.45);
  \node[n, below, text=figink2] at (32,-5.55) {8 B written to the doorbell};
  \node[n, anchor=north east] at (128,-5.42) {16 B, 4 B per word};
\end{tikzpicture}%
  \caption{\code{mlx5} send WQEs for an RDMA write, built from 16\,B
    segments (ctrl: control, AV: address vector, raddr: remote address)
    and fetched in 64\,B basic blocks. Only DC WQEs carry the AV
    (\S\ref{sec:boundary:transport}). An inline payload replaces the
    data segment's pointer with a length word and the payload. The
    bottom row expands the control segment; the doorbell store carries
  its first 8\,B.}
  \label{fig:wqe-segments}
\end{figure}

\parheading{WQE anatomy.}
An \code{mlx5} send WQE is a concatenation of 16-byte segments packed
into 64-byte basic blocks, shown in Figure~\ref{fig:wqe-segments} for
an RDMA write.
The payload can either be named by the data segment or packed directly
into the WQE, trading the NIC's DMA read of the payload for a larger WQE
(\S\ref{sec:micro:doorbell-issue-cost}). DC transports additionally
include an address-vector segment
to identify the dynamically connected target (DCT).

% \begin{stickynote}[Byte swapping]
%   NVSHMEM and DeepEP byte-swap with shift-and-mask macros or PTX \code{prmt}
%   sequences (\code{HTOBE32} and friends). CUDA offers the generic
%   \code{__byte_perm} intrinsic but no dedicated endianness
% conversion, so every
%   implementation carries its own helpers.
% \end{stickynote}

\parheading{Submission.}
Doorbell posting is a two-step process. First, the submitting thread
updates the doorbell record (\code{dbrec}) with the index of the next free
WQE basic block, which the NIC reads to get the number of posted WQEs
\cite{Mellanox-PRM-2020}. The thread then performs
an ordered 64-bit store to the doorbell register on the UAR
to notify the NIC of available work. The doorbell store carries the first
8 bytes of the WQE's control segment, which pack the WQE index and
the QP number.

The doorbell targets the UAR's send-doorbell/BlueFlame region. WQEs
can either be copied into the BlueFlame buffer via write-combining
stores, or DMA-fetched by the NIC separately. Inline copying inflates
the MMIO write payload of the doorbell, whereas NIC fetching preserves a
fixed doorbell size regardless of message length.
Every GPU-side implementation we evaluate opts for fixed-size
notifications and NIC-fetched WQEs
\cite{nvshmem-device-source,nccl-doca-qp-source,deepep-device-source},
whereas host-side \code{rdma-core} inline-copies small WQEs into the
BlueFlame buffer \cite{rdma-post-source}.

% \begin{stickynote}[Doorbell submission on the host]
%   \code{rdma-core} chooses the other way for small requests: it
%   copies the entire WQE into the BlueFlame buffer with write-combining stores,
%   so the MMIO write grows with the request up to the BlueFlame buffer size,
%   which is typically 256\,B. Past that size, or when \code{MLX5_SHUT_UP_BF}
%   forces a non-cached mapping, it falls back to the 8-byte doorbell
%   \cite{rdma-core}.
% \end{stickynote}

\parheading{Doorbell Batching.}
Announcing several WQEs with one doorbell raises throughput, with
a possible latency penalty if submitters wait to form batches.
The mechanism has been studied for CPU-side RDMA \cite{kalia-atc16},
and GPU libraries make use of the same optimization.
NVSHMEM defers the doorbell until
a configured number of WQEs is ready (default 32), but posts
isolated operations immediately
\cite{nvshmem-device-source,nvshmem-env-source}. NCCL GDAKI exposes
caller-controlled aggregation through its device API and DOCA posting
path \cite{nccl-gdaki-source,nccl-doca-qp-source}, and DeepEP's
low-latency kernels ring on every fourth message of an expert's queue
\cite{deepep-device-source}.
The batch a library achieves therefore depends on how many threads
post at once and in what order they publish, not only on its
threshold. We measure the effects of batching in
Section~\ref{sec:micro:concurrency}.

\parheading{Sharing a QP.}
Threads can share a QP by reserving consecutive WQE slots, building
their WQEs in parallel, and publishing them in reservation order before
one doorbell announces them. Publication is serialized: a thread cannot
publish past an unfinished predecessor, so lanes that drift apart wait
for each other. Cooperative publication avoids this by letting a warp
reserve, build, and publish its WQEs together, paying the reservation
and the doorbell once per warp
\cite{nvshmem-device-source,nccl-doca-qp-source}.
Section~\ref{sec:micro:concurrency} compares these strategies.

\parheading{Ordering.}
\label{sec:boundary:ordering}
Correct GPU-side RDMA delivery hinges on two ordering requirements:
(1) WQE and source-payload stores must be visible to the NIC before the
doorbell that announces them; and (2) when one thread rings the
doorbell for WQEs that other threads wrote, each writer's stores must
be visible before the ringing thread sees its slot as ready.
NVSHMEM implements both with plain stores by default, each preceded by
\code{__threadfence()} (\code{__threadfence_system()} when the queues
reside in host memory). When built without host-side queue
support\footnote{\code{NVSHMEM_IBGDA_SUPPORT_GPUMEM_ONLY=ON}}, it instead uses
GPU-scope release stores\footnote{PTX
\code{st.release.gpu.global.L1::no_allocate}}
on the doorbell path. NCCL's bundled DOCA implementation supports a
GPU-scope release
fence followed by a relaxed MMIO store, along with architecture and
scope-dependent alternatives \cite{nccl-doca-qp-source}. We measure
the cost of each of these orderings in
Section~\ref{sec:micro:doorbell-issue-cost}.

% \paragraph{Contrast with host-side RDMA}
% On the host side, \code{rdma-core} maps UARs write-combining (WC)
% by default and
% non-cached (NC) under \code{MLX5_SHUT_UP_BF}: WC enables BlueFlame
% but requires
% explicit barriers to flush the combining buffers, while NC disables
% BlueFlame and
% relaxes the flushing discipline \cite{rdma-core}. These knobs do
% not carry over
% verbatim to the GPU path. DEVX-allocated UARs fix their type at
% allocation time
% (\code{MLX5DV_UAR_ALLOC_TYPE_BF}, \code{NC}, or \code{NC_DEDICATED}); NVSHMEM
% prefers a dedicated non-cached UAR where \code{rdma-core} supports
% it and falls
% back to a BlueFlame-type allocation otherwise; and the mapping the
% GPU actually
% stores through is established by \code{cuMemHostRegister} as I/O memory. The
% ordering behavior of GPU MMIO stores over PCIe therefore differs from
% the CPU's WC
% discipline, and Section~\ref{sec:micro:doorbell} compares BF and NC-type UARs
% under identical GPU-side code.

\subsection{The completion path}
\label{sec:boundary:completion}
Upon executing a signaled WQE, the NIC
writes a CQE that identifies the
completed work and reports its status. Because send queues are strictly
ordered, a single CQE implicitly retires all preceding unsignaled WQEs
on the same queue \cite{rdma-post-send,rdma-post-source}. GPU threads poll
the CQ ring to consume these entries, but completion scope varies
across libraries. NVSHMEM's \code{quiet} walks all configured RC QPs
and the DCI pool, while DeepEP V1 polls a selected
\code{(peer, QP)}, usually one QP per local expert
\cite{nvshmem-device-source,deepep-device-source}.
NCCL GIN's \code{flushAsync} and \code{wait} complete one peer on
one context, while \code{flush} covers the context's pending transfers
for the participating threads \cite{nccl-gin-api}.
We measure the cost of the completion scope in
Section~\ref{sec:micro:completion}.

% \begin{stickynote}[Quiet semantics]
%   \code{nvshmem_quiet} polls the CQs of all DCI and RC QPs,
% optionally narrowed
%   by a PE or QP hint, up to their ready heads, and additionally issues a
%   consistency operation on a DCI if any \code{get} is outstanding.
% It guarantees
%   completion of prior operations at the initiator rather than their
%   visibility at
%   the target. DeepEP's vendored \code{nvshmemi_ibgda_quiet(dst_pe, qp_id)}
%   instead polls a single RC QP.
% \end{stickynote}

The NIC writes both remote payloads and local CQEs into GPU memory
by DMA, and an SM accessing them observes them within the GPU
memory hierarchy. Under its relaxed memory model, a thread that
observes a given CQE --- or any arbitrary flag in memory ---
is not thereby guaranteed to observe the payload written before it.
Libraries handle this in their wait and signal
primitives, and user code that polls raw memory must also handle it
for correctness. This is a common hazard for persistent kernels,
which lack the implicit synchronization of a kernel-launch boundary
\cite{rocshmem-ppopp,Landscape_Of_GPU_Comm,multi-gpu-iterative-solvers,
trotter2025cpu}, and for barrier-free NVLink collectives, which avoid
it by writing the flag and the data in one atomic store
\cite{sol-collectives}.

\subsection{Transport choice}
\label{sec:boundary:transport}
The transport decision influences both communication latency and
scalability. Reliable Connection (RC) transport binds each local QP
to one remote QP at creation time. Pre-establishing this connection
eliminates the need to specify destination addresses in individual
send WQEs, but forces each peer to establish a dedicated QP for
every other remote endpoint. With parallel submitters needed to
saturate the NIC (\S\ref{sec:micro:concurrency}), RC QPs may be allocated
per GPU, CTA, or warp, growing the local QP state. The associated QP
contexts, CQs, WQEs, and memory translations occupy NIC caches and
host-backed state, and poor locality across them reduces message rate
\cite{rdma-microarchitecture}. Section~\ref{sec:micro:icm} measures this
for GPU-driven traffic.

Dynamically Connected (DC) transport, by contrast, lets a pool
of dynamically connected initiators (DCIs) address many remote DC
targets (DCTs).
This reduces the number of initiator QPs at the cost of an additional
WQE segment in each send. Each DC send WQE carries an address vector
of 16\,B (48\,B on HCAs without the compact form~\cite{rdma-core})
identifying its destination. Changing a DCI's destination
may incur additional work on the NIC
\cite{nvshmem-device-source,nvidia-dc}. Section~\ref{sec:micro:dc}
measures both costs. The same per-peer state pressure has motivated
transport redesigns beyond InfiniBand: the Ultra Ethernet Transport
replaces the long-lived per-peer connection state of RC with
packet-delivery contexts that are established and released on demand
\cite{uec-spec}.

\section{Mechanism Evaluation}
\label{sec:Microbenchmarks}

We organize the experiments and evaluations around the three
questions introduced in
Section~\ref{sec:Introduction}.

\parheading{Minimal Implementations.} We implement two minimal paths
to accompany the production libraries listed in Table~\ref{tab:impls}.
\textbf{mini-gda} is a GPU-initiated
communication implementation with GPU-resident NIC queues
and the doorbell register mapped into GPU address space. GPU threads
construct WQEs, update queue state, ring the doorbell,
and poll completions, without a surrounding library. Ordering is
selected at build time, and payload placement, queue count, and doorbell
batching are launch parameters.
\textbf{mini-proxy} is a proxy implementation where GPU threads
write 16\,B request descriptors into rings in host memory, from which
CPU workers construct RDMA writes and submit them through \code{ibverbs}.
Worker count, ring count, and the number of requests submitted together
are configurable.
We measure DeepEP V1, whose device path uses NVSHMEM. DeepEP V2
replaces the NVSHMEM backend with NCCL GIN; since we measure GIN's
GDAKI path directly, the two cover both DeepEP backends at the
mechanism level.

\parheading{Platforms.} Table~\ref{tab:platforms} shows the platforms
used in our
evaluation. \textbf{P-IB} is the primary platform
for GPU-submitted networking and the proxy comparison.
\textbf{P-RoCE} is used for the SM-clock and completion-scope
experiments, as well as cross-platform controls. \textbf{P-H100} is a
larger cluster without
NIC doorbell mapping support used for active-connection experiments.
\textbf{P-GB200} is a Grace--Blackwell system with coherent memory
between the CPU and the GPU, where host rings and
counters used by proxies travel over NVLink-C2C
rather than PCIe. We repeat the one-operation and proxy
experiments on P-GB200 with one GPU and one NIC per node.
We use 8\,B operations to expose per-operation control costs, then
increase payload size to identify when link bandwidth dominates.

\parheading{Timing and Validation.} We time in-kernel operations with
\code{globaltimer} and use \code{clock64} to record the effective SM clock.
End-to-end measurements use CUDA events. We validate payloads after each
run and, where applicable, check NIC counters against the expected
traffic, including protocol overhead\append{
(Appendix~\ref{app:methodology})}.

\parheading{Proxy Clock State.}
On P-IB, CPU-proxy latency and rate depend on host operating state.
We compare \emph{cold} workers after 45\,s idle with \emph{warm}
workers after 20\,s of sustained load; telemetry shows base and turbo
clocks, respectively. P-IB proxy tables report cold values unless
marked\append{; Appendix~\ref{app:proxy-states} gives both states}.

\parheading{Pairs and Placement.} Unless stated otherwise, latencies are p50
over three independent process runs
on one pinned pair of nodes, using the NIC attached to
the GPU's PCIe switch.

\begin{table}[!tbp]
  \centering
  \caption{Evaluation platforms.}
  \label{tab:platforms}
  \setlength{\tabcolsep}{2pt}
  \begin{tabularx}{\columnwidth}{@{}l Y Y Y Y@{}}
    \toprule
    & \textbf{P-IB} & \textbf{P-RoCE} & \textbf{P-H100} & \textbf{P-GB200} \\
    \midrule
    Nodes $\times$ GPUs   & $4 \times 4$ H200 & $2 \times 8$ B200 &
    $100 \times 4$ H100 & $2 \times 1$ GB200 \\
    CPU & Xeon 6548Y+ & Xeon 8581C & Xeon 8460Y+ & Grace \\
    NICs / node           & $4\times$ CX-7 & $8\times$ CX-7 & $4\times$ CX-7
    & $1\times$ CX-7 \\
    Link                  & 200\,Gb/s IB & 400\,Gb/s RoCEv2 & 200\,Gb/s IB
    & 400\,Gb/s IB \\
    Driver / CUDA         & 580.95 / 13.0 & 590.48 / 13.2 & 595.71 / 12.6
    & 580.159 / 13.0 \\
    NVSHMEM               & 3.4.5 / 3.7.2 & 3.7.2 & 3.4.5 / 3.7.2 & 3.7.2 \\
    NCCL                  & 2.30.7 / 2.31.2 & 2.31.2 & --- & 2.31.2 \\
    \midrule
    Doorbell writer       & GPU & GPU & host thread & GPU \\
    \bottomrule
  \end{tabularx}
\end{table}

% ============================================================================
\subsection{The cost of one operation}
\label{sec:micro:doorbell-issue-cost}

\parheading{Question.} What does one GPU-initiated RDMA write cost the issuing
thread, how much of that is the mechanism and how much the library, and how
does handing the operation to a CPU proxy compare?

\parheading{Setup.}
One GPU thread issues 8\,B puts to a remote processing element (PE),
with at most one outstanding \code{put}. We time the following:
\begin{itemize}
  \item \textbf{Issue:} Time the GPU thread spends submitting
    the \code{put}, including WQE construction, queue management,
    ordering, and doorbell stores. Completion time is excluded.
  \item \textbf{Put+completion:} Issue and completion-routine
    return, including polling, ordering, and queue updates.
  \item \textbf{Round trip:} Time to send a request and observe the remote
    GPU's reply.
\end{itemize}

We compare three GPU-submitted stacks:
\begin{itemize}
  \item \textbf{mini-gda:} One RC QP, GPU-scope ordering as in DOCA,
    and completion by polling its CQ.
  \item \textbf{NCCL GIN GDAKI:} One context and one RC QP per peer,
    completing with \code{flushAsync} and \code{wait} without
    doorbell aggregation.
  \item \textbf{NVSHMEM:} Two RC QPs per peer (default) through the internal
    \code{nvshmemi_ibgda_rma_nbi} + internal quiet, and public
    \code{nvshmem_putmem_nbi} + \code{nvshmem_quiet}. We separately measure
    16 QPs, which achieves the highest message rate
    (\S\ref{sec:micro:concurrency}).
\end{itemize}

In addition to the GPU-submitted paths, we report NVSHMEM IBRC, NCCL GIN
Proxy, and mini-proxy as CPU-submitted references. A proxy's issue interval
measures GPU enqueue without the host-side WQE construction. Round trips
use each stack's own notification mechanism: put-with-signal for NVSHMEM,
\code{putValue} with a signal increment for NCCL, and a polled data word
for mini-gda and mini-proxy.

\begin{table}[t]
  % figure source on google docs
  \centering
  \caption{One 8\,B operation. Completion is one CQE for
    mini-gda, a host-resident counter for the baseline mini-proxy, and
  a GPU-resident counter written through GDRCopy for the tuned one.}
  \label{tab:oneop}

  \setlength{\tabcolsep}{2.6pt} % Fit within the USENIX column as well.
  \begin{tabular}{@{}l S[table-format=1.2] S[table-format=2.2]
      S[table-format=2.2]
      @{\hspace{1em}} S[table-format=1.2] S[table-format=2.2]
    S[table-format=2.2]@{}}
    \toprule
    & \multicolumn{3}{c}{\textbf{P-IB} (\si{\micro\second})} &
    \multicolumn{3}{c@{}}{\textbf{P-GB200} (\si{\micro\second})} \\
    \cmidrule(lr){2-4} \cmidrule(l){5-7}
    \textbf{Path} & {\textbf{Issue}} & {\textbf{Put+c.}} & {\textbf{RTT}}
    & {\textbf{Issue}} & {\textbf{Put+c.}} & {\textbf{RTT}} \\
    \midrule
    \tabgroup{7}{GPU submits}
    mini-gda, inline & 0.70 & 4.03 & 6.85 & 0.77 & 5.70 & 8.64 \\
    mini-gda, non-inline & 0.70 & 4.64 & {---} & 0.80 & 6.62 & {---} \\
    mini-gda, unordered (unsafe) & 0.19 & 3.46 & {---} & 0.16 & 5.09 & {---} \\
    GDAKI, inline & 1.60 & 6.02 & 10.50 & 1.60 & 7.26 & 14.40 \\
    GDAKI, non-inline & 1.82 & 6.85 & {---} & 1.79 & 8.48 & {---} \\
    NVSHMEM internal & 4.26 & 9.18 & {---} & 5.50 & 12.19 & {---} \\
    NVSHMEM public & 5.31 & 11.10 & 21.50 & 6.62 & 13.82 & 25.25 \\
    \midrule
    \tabgroup{7}{CPU proxy submits}
    NVSHMEM IBRC & 0.99 & 6.40 & 11.07 & 1.18 & 7.14 & 12.42 \\
    NCCL GIN Proxy & 2.46 & 8.19 & 16.32 & 2.27 & 7.87 & 17.54 \\
    mini-proxy, baseline & 2.27 & 7.07 & 14.59 & 2.24 & 6.91 & 14.91 \\
    mini-proxy, tuned & 0.13 & 4.10 & 5.89 & 0.13 & 4.99 & 7.46 \\
    \bottomrule
  \end{tabular}
\end{table}

\parheading{The mechanism costs \SI{0.7}{\micro\second} to issue.}
Constructing the WQE, updating the doorbell record, and writing the doorbell to
the NIC takes mini-gda \SI{0.70}{\micro\second}
(Table~\ref{tab:oneop}).
GDAKI is the next fastest GPU-submitted path, and NVSHMEM's internal
and public paths add significant further costs before completion
(Table~\ref{tab:oneop}). Replacing NVSHMEM's default fence with a
GPU-scope release recovers only \SI{0.16}{\micro\second} of issue on the same
pair. NVSHMEM's additional work includes slot reservations, ready-head
updates, doorbell locks, and QP lookups
\cite{nvshmem-device-source,nccl-doca-qp-source}. The SM-clock experiment
later in this section tests how much of the issue gap scales with
device execution speed. P-GB200 repeats the ranking of the
GPU-submitted paths (Table~\ref{tab:oneop}).

\parheading{Ordering changes the latency floor.}
WQE stores must become visible before the doorbell announces them
for the NIC to read the correct data (\S\ref{sec:boundary:ordering}).
Table~\ref{tab:ordering-short} compares different ordering
configurations used by production libraries in mini-gda.
GPU-scope release ordering has a lower cost than system scope on both
platforms: a system-scope fence costs $3.7\times$ the issue time of a
GPU-scope fence on P-IB (2.62 vs.\ \SI{0.70}{\micro\second}).
Shen et al.\ report the same pattern
over NVLink, where one barrier costs more than \SI{1}{\micro\second} against a
\SI{1.4}{\micro\second} data-movement floor \cite{sol-collectives}.
Omitting fences and using plain CQ loads achieves the lowest
latencies --- \SI{0.19}{\micro\second} issue and
\SI{3.46}{\micro\second} through completion --- but
is not guaranteed to be safe (\S\ref{sec:boundary:ordering}).
We found no corruption with the ordering removed. While this does not prove
it safe in general, an implementation could drop the fence for a
workload it has verified to get closer to the mechanism's floor.

\begin{table}[t]
  \centering
  \caption{mini-gda ordering configurations. Each row orders
    the WQE and doorbell-record stores before the doorbell store
    (\S\ref{sec:boundary:ordering}) and polls the CQ at the matching
    scope.
  }
  \label{tab:ordering-short}
  \begin{tabular}{@{}l S[table-format=1.2] S[table-format=1.2]
    @{\hspace{1em}} S[table-format=1.2] S[table-format=1.2]@{}}
    \toprule
    & \multicolumn{2}{c}{\textbf{P-IB} (\si{\micro\second})} &
    \multicolumn{2}{c@{}}{\textbf{P-GB200} (\si{\micro\second})} \\
    \cmidrule(lr){2-3} \cmidrule(l){4-5}
    \textbf{Ordering} & {\textbf{Issue}} & {\textbf{Put+c.}}
    & {\textbf{Issue}} & {\textbf{Put+c.}} \\
    \midrule
    None (unsafe control) & 0.19 & 3.46 & 0.16 & 5.09 \\
    GPU-scope fence (DOCA, GDAKI) & 0.70 & 4.03 & 0.77 & 5.70 \\
    \quad + fenced doorbell record & 0.93 & 4.26 & 0.99 & 5.89 \\
    GPU-scope release store (DeepEP) & 0.70 & 4.00 & 0.77 & 5.66 \\
    \code{__threadfence()} (NVSHMEM) & 1.22 & 4.48 & 1.22 & 6.14 \\
    System-scope release store & 2.37 & 5.66 & 2.08 & 7.01 \\
    System-scope fence & 2.62 & 5.92 & 2.34 & 7.26 \\
    \bottomrule
  \end{tabular}
\end{table}

\parheading{Completion arrives \SI{3}{\micro\second} after the doorbell.}
An instrumented run observes a valid CQE \SI{3.30}{\micro\second} after the
doorbell store, covering doorbell delivery, NIC and network processing,
and GPU polling. The rest of each path's put+completion time in
Table~\ref{tab:oneop} is spent in its completion routine, which bundles
polling, ordering, and queue updates, so we do not attribute it to one
mechanism.
Every GPU-submitted path completes 1.2--\SI{3.0}{\micro\second} later
on P-GB200 than on P-IB, while issue times stay within
\SI{1.3}{\micro\second}. On P-GB200 the NIC reaches GPU memory
through the Grace CPU and NVLink-C2C rather than through a shared PCIe
switch, so every WQE fetch, payload read, and CQE write takes a longer
path. Consistent with this, moving mini-gda's queues to host memory
there shortens put+completion by \SI{0.86}{\micro\second} at the same
ordering.

\begin{figure}[t]
  \centering
  \includegraphics[width=\columnwidth]{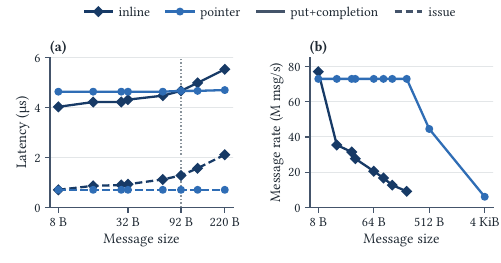}
  \caption{Inline vs.\ pointer payloads in mini-gda on P-IB.
    (a)~One-thread issue and put+completion latency against inline
  payload size. (b)~Message rate at 16 QPs with 16 WQEs per doorbell.}
  \label{fig:e8-inline}
\end{figure}

\parheading{Inlining trades a source read for GPU stores.}
Inlining an 8\,B value spares the NIC a read of the source buffer
at no extra issue cost, saving \SI{0.61}{\micro\second} of completion
time. At larger inline payloads, however, the
issue cost increases, and the batched message rate falls
(Figure~\ref{fig:e8-inline}). We find that issue time grows by
about \SI{0.1}{\micro\second} per additional 16\,B chunk, and the completion
savings are gone by 92\,B. NVSHMEM, GDAKI, and DeepEP inline only 8\,B values.

\begin{figure}[t]
  \centering
  \includegraphics[width=\columnwidth]{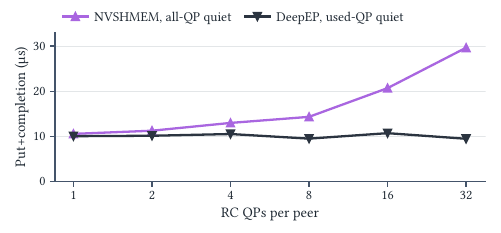}
  \caption{Completion scope against queue count on P-IB. Both quiet
    curves use the same internal put (NVSHMEM 3.4.5); the all-QP
    \code{quiet} sweeps every configured RC QP, while DeepEP's ported
  quiet polls only the used QP.}
  \label{fig:quiet-tax}
\end{figure}

\parheading{All-QP quiet grows with queue count.}
\label{sec:micro:completion}
While adding queues increases posting capacity, they each
individually require completions. NVSHMEM's PE-wide \code{quiet}
visits all configured QPs regardless of use, so its put+completion
latency doubles between 1 and 16 QPs (Figure~\ref{fig:quiet-tax}),
while DeepEP's ported single-QP quiet stays flat, as does a
signal-based round trip. NVSHMEM's own per-QP quiet is within
\SI{0.5}{\micro\second} of
the port at 16 QPs, so the gain comes from narrowing the completion
scope rather than faster queue handling.
NVSHMEM 3.7.2 on P-RoCE shows the same growth to 32
QPs\append{ (Appendix Table~\ref{tab:proce-scope})}.
Letting NVSHMEM size its DCI pool automatically adds a
further \SI{81}{\micro\second} on P-GB200 without changing issue
time\append{ (Appendix Table~\ref{tab:gb200-scope})}.

\begin{figure}[t]
  \centering
  % alternative colorings: fig02_clock_stretch_ramp.pdf (IBGDA
  % arms in a purple ramp), fig02_clock_stretch_metric.pdf (color by
  % measurement, marker by implementation).
  \includegraphics[width=\columnwidth]{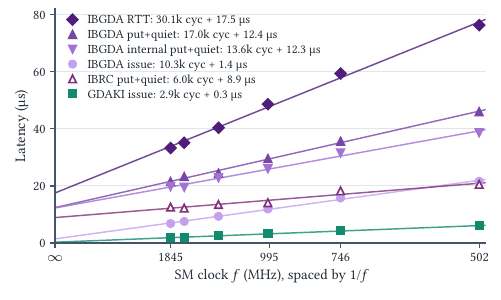}
  \caption{Latency against $1/f$, where $f$ is the locked SM clock.
    Both GPUs are on the same node, each with its own NIC across the fabric.
    The slope is an effective clock-sensitive coefficient in cycles and
  the intercept is the fitted clock-insensitive component.}
  \label{fig:clock-stretch}
\end{figure}

\parheading{A throttled SM is a slower communicator.}
On P-RoCE, lowering the SM clock slows issue more than put+completion
for both NVSHMEM and GDAKI (Figure~\ref{fig:clock-stretch}).
At the top clock, the clock-sensitive term is 80\% of NVSHMEM's
public issue and 85\% of GDAKI's, but NVSHMEM's coefficient is
$3.5\times$ larger, which accounts for most of the issue-latency gap.
A throttled GPU can therefore communicate more slowly over the same
network. Offloading to a CPU proxy does not remove this dependence, as the GPU
still enqueues the request: the fit attributes 92\% of IBRC's enqueue
time and 27\% of its put+completion
to the SM clock\append{ (Appendix Table~\ref{tab:clockfit})}.

% \begin{figure}[t] \centering
%   \includegraphics[width=\columnwidth]{Figures/Experiments/fig_proxy_queues_column.pdf}
% ... same caption ... \end{figure}
\begin{figure*}[t]
  \centering
  \includegraphics[width=\textwidth]{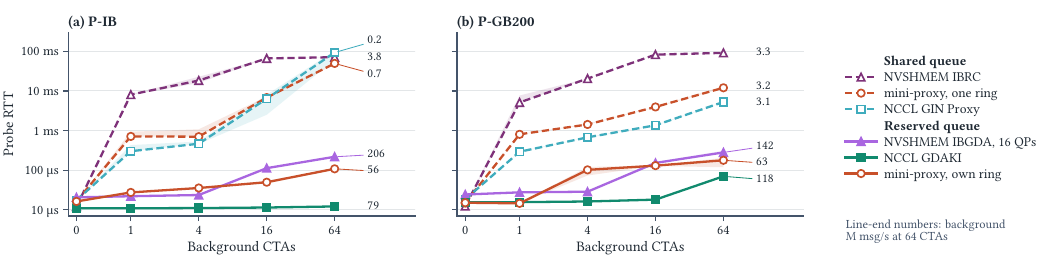}
  \caption{RTT under load on (a)~P-IB and (b)~P-GB200, with background
    CTAs loading both endpoints. Dashed curves share one ring or context
    with the bulk traffic; solid curves reserve a probe queue. IBGDA uses
    a fixed 16-QP pool and GDAKI a private context. Labels give the
  background M\,msg/s carried at 64 CTAs.}
  \label{fig:latload}
\end{figure*}

\parheading{Reducing the Proxy Enqueue Cost.}
mini-proxy's baseline protocol takes \SI{2.27}{\micro\second} to enqueue a
request, including a device atomic for ring reservation, a PCIe read
of host-resident progress, and a system-scope release. We can tune the proxy
to the workload by allocating one
producer per ring, caching progress, and encoding validity in the descriptor,
which together reduce enqueue to a single posted 16\,B store taking
\SI{0.13}{\micro\second}.
The tuned proxy matches mini-gda through completion on P-IB and
has a lower median round trip on both platforms
(Table~\ref{tab:oneop}). These results require a dedicated CPU
worker pinned to a core on the NIC's NUMA node
(\S\ref{sec:micro:proxydesign}).

\parheading{Proxy latency depends on host operating state.}
On P-IB, CPU-only warm-up improves proxy latency and tails
while leaving GPU-submitted paths largely unchanged. When warm, the
tuned proxy's round-trip p99 falls from 8.9 to \SI{5.6}{\micro\second},
and IBRC's small-message rate doubles from 1.9 to
3.8\,M\,msg/s\append{ (Appendix~\ref{app:proxy-states})}.
This sensitivity shows that worker placement and host-side
preconditioning must be considered for reproducible proxy comparisons.

\parheading{Takeaway.}
Software choices, not the hardware mechanism, set single-operation
latency. A minimal GPU path issues an 8\,B write in
\SI{0.7}{\micro\second} and completes in \SI{4.0}{\micro\second},
while libraries add up to \SI{4.6}{\micro\second} of issue time through
queue management, ordering, and completion scope. Ordering scope alone
changes issue time $3.7\times$ across safe configurations, and all-QP
completion doubles put+completion from 1 to 16 QPs while per-QP
completion stays flat. Issue time follows the SM clock on both paths,
including a proxy's GPU-side enqueue. A tuned proxy matches the
minimal GPU path through completion and has a lower median round
trip, at the cost of a dedicated core whose clock state sets its
latency, tail, and rate.

\subsection{The proxy design space}
\label{sec:micro:rtt-load}
\label{sec:micro:proxydesign}

\parheading{Question.}
Several libraries argue that a well-designed proxy matches GPU submission
\cite{uccl-ep,msccl++,pplx-garden-fabric-lib}, while others
\cite{NVIDIA-Blog-IBGDA,gpuinitiatednetworkingnccl,deepep2025} report
latency and message-rate wins for GPU submission. The libraries make
different choices for threads, rings, batching, and the GPU--CPU handoff
(Table~\ref{tab:impls}). When can a CPU proxy match GPU submission, and
which of these choices matter where?

\parheading{Setup.}
Four design choices differ across proxy implementations
(Table~\ref{tab:impls}): the descriptor rings ($R$); the CPU workers
($T$) that poll rings and post requests, each with its own NIC QP; the
batch of work requests ($B$) chained into one send (\code{ibv_post_send});
and the GPU--CPU handoff, that is, how GPU threads reserve slots, publish
descriptors, and observe worker progress.

We measure the round trip of
an 8\,B request-reply with and without background CTAs issuing writes
through the same NIC from another stream. We also report the message rate and
goodput when many GPU threads issue puts before waiting for completion.
Both endpoints generate background traffic. We report the rate carried
during the probe alongside latency, since each configuration
sustains a different total load.

Proxy workers are placed near the GPU's NIC: mini-proxy and
UCCL-EP pin workers to cores, while MSCCL++ binds them to the local NUMA
node. The loaded, message-rate, and payload experiments are repeated on P-GB200
for mini-proxy, IBRC, GIN Proxy, IBGDA, and GDAKI.
fabric-lib is host-initiated by design \cite{pplx-garden-fabric-lib}, so we
drive its native client from a host loop and time it on the CPU; its
GPU-to-host handoff is not measured.

\parheading{Latency at Idle.}
Published evaluations place proxies and GPU-submitted transports in overlapping
ranges, and Hamidouche et al.\ measure IBRC's round trip below
IBGDA's \cite{gpuinitiatednetworkingnccl}.
We reproduce this with IBRC
answering a round trip in \SI{11.1}{\micro\second} against NVSHMEM's public 21.5
(Table~\ref{tab:oneop}).
These results do not establish an intrinsic proxy advantage, as
the minimal GPU path and GDAKI both answer faster than IBRC.
The other proxies' round trips reflect their notification protocols:
fabric-lib answers in \SI{10.1}{\micro\second} with its write-immediate counter,
UCCL-EP in 13.1 polling the received word, and MSCCL++ in 18.5
with a semaphore and per-iteration flush.

\parheading{Shared queues suffer under load.}
The latency measurements shift when background traffic is introduced
(Figure~\ref{fig:latload}). Every proxy that sends the probe through
a queue shared with bulk traffic loses three orders of magnitude:
IBRC's single FIFO, mini-proxy with one ring, and GIN Proxy's shared
context all reach tens of milliseconds at 64 background CTAs while
carrying only 2--4\,M\,msg/s. IBRC, which reported competitive
round-trip latency at idle, collapses $670\times$ with a
single background CTA. These configurations force latency-sensitive
requests through the same FIFO as bulk traffic with no isolation or
priority. GPU submission does not isolate by itself either: a single GDAKI
context shared by the probe and all background CTAs answers in
\SI{362}{\micro\second} at 64 CTAs while carrying 2.6\,M\,msg/s,
whereas a private
context keeps p50 at 12--\SI{16}{\micro\second} with the background rate at
73--79\,M\,msg/s.

\parheading{Queue reservation reduces interference.}
Reserving a queue for the probe recovers most of the loss:
UCCL-EP and MSCCL++ go from milliseconds to 120 and \SI{825}{\micro\second}.
With workers, rings, and batching held fixed, a private mini-proxy
ring cuts the loaded round trip by one to two orders of magnitude on
both platforms, and a private UCCL-EP FIFO or a reserved MSCCL++
service does the same (Figure~\ref{fig:latload}\append{,
Appendix~\ref{app:proxy-details}}).
GPU submission exhibits the same behavior:
with GDAKI's 65-context pool and carried load held fixed, moving the
probe off a context shared with a single background CTA halves its
latency, from 27.3 to \SI{13.9}{\micro\second}\append{ (Appendix
Table~\ref{tab:proxy-controls})}.
Reserving a worker as well adds little and takes capacity from bulk
traffic, but shared workers and NIC resources still permit
interference, so
latency must also be read with the load carried. Paced to the
same 30\,M\,msg/s, a reserved GDAKI context still answers in
less than half the time of a reserved mini-proxy ring, and
comparing them at their unrestricted loads would mix isolation
with posting capacity.

\begin{figure}[t]
  \centering
  \includegraphics[width=\columnwidth]{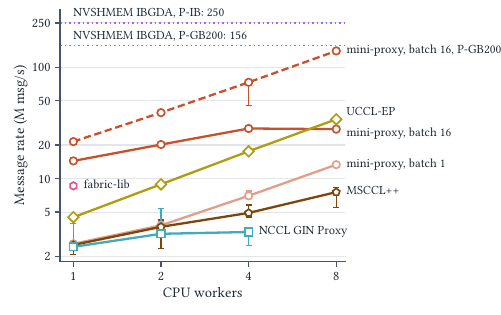}
  \caption{8\,B proxy message rate against worker count on P-IB.
    Batch counts the requests mini-proxy chains into one post; batch 16
    is repeated on P-GB200 (dashed).
    fabric-lib runs one worker per NIC. Dotted lines mark NVSHMEM
    IBGDA on the same pairs. Worker cores are at their base clock;
    turbo raises every proxy rate 1.3--$3.2\times$\append{
  (Appendix~\ref{app:proxy-states})}.}
  \label{fig:proxy-capacity}
\end{figure}

\begin{figure}[tbp]
  \centering
  \includegraphics[width=\columnwidth]{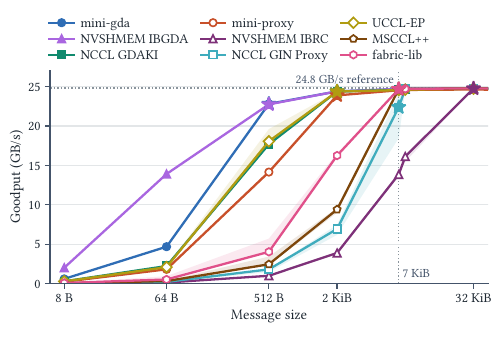}
  \caption{Pipelined goodput against payload size on P-IB. Stars mark
    the first sampled size at 90\% of the 24.8\,GB/s reference.
    Vertical line is a DeepSeek-V3 FP8 token vector (7,168\,B). Proxies
    use four workers (UCCL-EP eight, fabric-lib and IBRC one). IBRC is
    measured cold and the other proxies' clock state was not
  controlled. Producer CTAs fall from 16 to one with size.}
  \label{fig:crossover-proxies}
\end{figure}

\parheading{Message rate: chaining and workers raise capacity.}
As with GPU submission, proxy message rate can be raised by batching
(chaining) and parallel submitters.
GPU threads construct WQEs in parallel, whereas each CPU worker
serializes its posting work.
Chaining requests amortizes that work, and additional workers
provide parallel submission (Table~\ref{tab:impls}).
Libraries implement a mix of the two: MSCCL++\footnote{We patched a
  shared-pointer refcount bug that serializes
  MSCCL++'s services\append{ (Appendix~\ref{app:msccl-artifact})};
  we use this build throughout and retain the one-request posting
policy.} adds
workers as services but posts each write individually, fabric-lib chains up to
four requests on one worker, UCCL-EP drains several FIFOs into
variable-length chains, and mini-proxy exposes $T$, $R$, and $B$
separately\append{ (Appendix Table~\ref{tab:proxy-capacity-details})}.
Figure~\ref{fig:proxy-capacity} shows the resulting scaling.
At $T=1$, $R=32$, increasing mini-proxy's batch limit $B$ from 1 to 16
raises the message rate $5.5\times$. Adding workers scales every library
tested up to a point, though they still stay an order of magnitude
below NVSHMEM IBGDA's 250\,M\,msg/s (\S\ref{sec:micro:concurrency}).
Under load,
the batch costs little latency: with one worker, eight rings, and 64
background CTAs, a private probe answers in \SI{272}{\micro\second} at $B=16$
against \SI{255}{\micro\second}
at $B=1$ while the background carries $5.2\times$ more traffic.
Independent workers still need to avoid contending on shared state:
GIN Proxy stays near 3\,M\,msg/s over one to four workers in the
tested four-context configuration, and MSCCL++'s services scale only
with the refcount patch. fabric-lib's single worker reaches
8.6\,M\,msg/s by amortizing its native calls across paged chains of
up to four requests.

The proxy's ceiling is platform-dependent. On P-GB200, mini-proxy
continues scaling to eight workers at $B=16$, reaching
140\,M\,msg/s, or 90\% of IBGDA's 156\,M\,msg/s there, against 250
on P-IB (Figure~\ref{fig:proxy-capacity}), although CPU, link,
firmware, and RDMA provider all differ between the pairs. Coherent
NVLink-C2C helps capacity but not latency: the tuned proxy's round trip
is \SI{7.5}{\micro\second} on P-GB200 against \SI{5.9}{\micro\second}
on P-IB (Table~\ref{tab:oneop}).

\parheading{Bandwidth: larger payloads hide the posting-rate differences.}
Posting rate sets the payload size needed to fill the link
(Figure~\ref{fig:crossover-proxies}). The experiment compares three
GPU-submitted paths with six proxy implementations on one pair.
The first measured size reaching 90\% of 24.8\,GB/s is 512\,B for
NVSHMEM IBGDA and mini-gda. GDAKI, mini-proxy, and UCCL-EP reach it
at 2\,KiB, while MSCCL++, GIN Proxy, and fabric-lib lag until 7\,KiB.
Paths with higher small-message rates saturate the link with smaller payloads.
At 7,168\,B, the size of a DeepSeek-V3 token's FP8 hidden-state
vector \cite{deepseek-v3-config}, all paths in the sweep except cold
IBRC exceed 90\%
of the reference bandwidth; IBRC reaches it at 32\,KiB cold and
7\,KiB warm.
On P-GB200's 400\,Gb/s link, mini-proxy,
IBGDA, and GIN Proxy also approach line rate at 4\,KiB\append{
(Appendix Table~\ref{tab:gb200-payload})}.
Completion latency remains different even as goodput
converges\append{ (Appendix~\ref{app:payload-latency})}.

\parheading{Takeaway.}
Idle latency reflects per-operation software costs, while loaded latency
depends on queue isolation and carried traffic. Separate rings reduce
interference, while workers and chaining raise proxy capacity.
Larger payloads hide posting-rate differences in pipelined goodput,
but completion latency remains a separate consideration. Proxy
capacity stays well below GPU submission on P-IB but reaches 90\% of
IBGDA's rate on P-GB200.

\subsection{Message rate and resource costs}
\label{sec:micro:cost}

\subsubsection{\textbf{What does message rate take?}}
\label{sec:micro:concurrency}
\label{sec:micro:transport}
\label{sec:micro:mapping}
\label{sec:micro:submission}

% \parheading{Question.}
A single submitting thread cannot saturate the NIC, and concurrent
submission is needed.
How does message rate grow with threads, CTAs, and QPs, and what does
doorbell batching add?

\parheading{Setup.}
On P-IB, GPU threads issue 8\,B inline puts to one remote GPU
with many writes outstanding, periodic queue reclamation, and a final
drain.
mini-gda varies threads, QPs, SMs, and WQEs announced
per doorbell (Table~\ref{tab:ladder}).
NVSHMEM runs with 16 RC QPs and 256 threads per CTA, and GDAKI with one
context and eight threads per CTA, each library's best measured
configuration.

\begin{table}[t]
  \centering
  \caption{mini-gda 8\,B message rate on P-IB. Per QP
    is the total rate divided by the QP count. $\ddagger$: one QP per CTA shared
    cooperatively, reserved per warp and rung once per 32 WQEs, as NVSHMEM does.
    $\dagger$: one doorbell lock per WQE. $^\ast$: slots handed off
    per thread rather than per warp.
  }
  \label{tab:ladder}
  \footnotesize
  \setlength{\tabcolsep}{3pt}
  \renewcommand{\arraystretch}{1.05}
  \begin{tabular}{@{}c r r r r r r@{}}
    \toprule
    & \textbf{Thr.} & \textbf{QPs} & \textbf{SMs} & \textbf{Batch} &
    \textbf{M\,msg/s} & \textbf{per QP} \\
    \midrule
    \band{3}{One\\thread}
    & 1              & 1     & 1  & 1  & 1.79 & 1.79 \\
    & 1              & 2--32 & 1  & 1  & 1.85 & ---  \\
    & 1              & 1     & 1  & 16 & 4.70 & 4.70 \\
    \cmidrule{1-7}
    \band{4}{One\\QP}
    & 32$^\dagger$   & 1     & 1  & 1  & 1.27 & 1.27 \\
    & 256$^\ast$            & 1     & 1  & 32 & 2.60 & 2.60 \\
    & 32$^\ddagger$  & 1     & 1  & 32 & 21.6 & 21.6 \\
    & 256$^\ddagger$ & 1     & 1  & 32 & 30.7 & 30.7 \\
    \cmidrule{1-7}
    \band{5}{Several\\QPs}
    & 16             & 16    & 16 & 1  & 27.9 & 1.74 \\
    & 16             & 16    & 16 & 16 & 74.4 & 4.65 \\
    & 32             & 32    & 1  & 16 & 133  & 4.15 \\
    & 64             & 64    & 1  & 16 & 206  & 3.21 \\
    & 64             & 64    & 64 & 16 & 260  & 4.06 \\
    \midrule
    \band{2}{NVSHMEM\\v3.7.2}
    & 256            & 1     & 1  & 32 & 25.2 & 25.2 \\
    & 4,096          & 16    & 16 & 32 & 250  & 15.6 \\
    \bottomrule
  \end{tabular}
\end{table}

\parheading{Cooperative sharing raises the rate of one queue.}
One thread with one QP and one WQE per doorbell posts
1.79\,M\,msg/s (Table~\ref{tab:ladder}).
Adding parallelism to either the QPs or the threads individually does
not meaningfully change the posting rate.
Sharing the QP cooperatively instead
raises the rate: a warp can reserve 32 slots and build its WQEs
in lockstep, with one lane publishing the group with a single ring
(as NVSHMEM does). The same QP in this regime posts 21.6\,M\,msg/s
from one warp and 30.7 from two or more warps, because the
reservation, the ordered doorbell sequence, and the QP lock are paid
once per warp rather than once per WQE. Adding more threads or larger
doorbell batches does not raise the single-QP ceiling in our experiment.

\parheading{Batching raises the per-thread rate and queues multiply it.}
Batching amortizes the ordered doorbell sequence, including its 8\,B
MMIO store, across several WQEs. Announcing 16 WQEs per
doorbell gets one thread a $2.6\times$ boost from 1.79 to
4.70\,M\,msg/s, with diminishing gains from larger batches.
GDAKI's doorbell aggregation raises its rate
by about $6\times$ per context, and NVSHMEM
batches ready WQEs across threads. This is the GPU equivalent of
host-side doorbell batching \cite{kalia-atc16}.

\parheading{GPU submission is not needed for peak rate.}
A host thread ringing the doorbells for GPU-built WQEs reaches
257\,M\,msg/s at 16 active QPs on P-H100 and 258\,M\,msg/s on P-IB,
at the cost of 2.6--\SI{5.8}{\micro\second} more put+completion.

\parheading{Publication order limits library rates.}
The libraries fill their queues differently. NVSHMEM batches ready
WQEs across warps; GDAKI aggregates per thread but publishes in
reservation order. Lanes that drift apart therefore wait for each
other. At 32 threads per context, a \code{__syncwarp} after each put
roughly triples GDAKI's rate\append{ (Appendix~\ref{app:publication})}.
The minimal path likewise shows that cooperative publication is
much faster than handing off individual slots on the same QP.
More submitting threads help only when publication can keep up.

\parheading{Takeaway.}
Message rate depends on independent queues, batching, and how warps
and QPs are shared: mini-gda and NVSHMEM reach the 250--260\,M\,msg/s
ceiling with 16 shared QPs and 4,096 threads. mini-gda also reaches
this rate with
64 batched single-thread QPs. Increasing NVSHMEM from the default 2 to 16
QPs raises rate about $5\times$ at one CTA per QP.
The queues that buy rate are paid for elsewhere: each
lengthens a PE-wide quiet (\S\ref{sec:micro:completion}) and is one more
active connection for the NIC (\S\ref{sec:micro:icm}).

\subsubsection{\textbf{What does the enclosing kernel pay?}}
\label{sec:micro:wqecost}
\label{sec:micro:deepep}

% \parheading{Question.}
Communication code can change a kernel's
register allocation, block residency, and spilling even when unused.
Section~\ref{sec:micro:doorbell-issue-cost} measured the submitting
thread's latency; what does carrying this code,
executing it, and waiting for completion cost the kernel's useful work?

\parheading{Setup.} On P-IB, compute (FMA) and memory-bound streaming
kernels keep $N$ live values per thread across communication calls, in
528 blocks of 256 threads (four per SM on average), so every variant has
the same useful work and launch shape. Each block sends 0, 1, 4, or 16
individual 8\,B or 7\,KiB messages spread through its work. We compare a
kernel without communication code to three variants: \textbf{code only},
where the path is compiled in but its branch is never taken; \textbf{send,
then compute}, where one thread per block submits each message and its
warp resumes computing while the transfer proceeds; and \textbf{send, wait,
then compute}, where that warp first waits for the message's completion.
Other warps compute throughout.
mini-gda (one QP per block) and
mini-proxy (four workers, 32 rings) are the bare paths; NVSHMEM (16 RC
QPs per PE) is built both with separately compiled device functions and
with its body inlined into the caller, and GDAKI (32 contexts)
submits with \code{put} plus \code{flushAsync}. Times measure kernel
execution, excluding the final completion drain\append{
(Appendix~\ref{app:kernel-details})}.

\begin{figure}[t]
  \centering
  \includegraphics[width=\columnwidth]{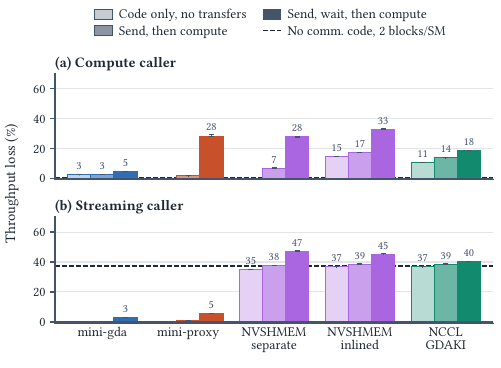}
  \caption{Useful-work throughput loss relative to the
    kernel without communication code, for (a)~a compute caller with
    eight live values (2.1\,ms) and (b)~a streaming caller with 16 live
  values (2.6\,ms). The three bars of an arm are not additive.}
  \label{fig:kernelcost}
\end{figure}

\parheading{Dormant communication code can reduce throughput.}
Adding communication increases register use, but its throughput cost
depends on the caller. In the compute caller
(Figure~\ref{fig:kernelcost}a), mini-gda, mini-proxy, and separately
compiled NVSHMEM retain nearly all
throughput, while inlined NVSHMEM and GDAKI lose 11--15\%. In the
streaming caller, both NVSHMEM builds
and GDAKI lose about 37\% before sending a message, consistent with
lower block residency (Figure~\ref{fig:kernelcost}b).

A separate run with eight blocks per SM shows that even mini-gda's
dormant path costs a streaming caller 27\% by lowering residency.
Inlining also changes the trade-off: it raises register demand in a
thin caller but postpones spilling as the caller's live state grows,
because a separately compiled call forces the caller's live values to
survive it\append{ (Appendix~\ref{app:kernel-details})}.

\parheading{Waiting costs more than intermittent submission.}
Sending without waiting adds little loss, since even at 16 messages per
block the compute caller averages only about 4\,M\,msg/s. Waiting for
each 8\,B message reduces throughput by a further 2--5\% with
mini-gda or GDAKI and
19--27\% with NVSHMEM or mini-proxy. Queue count amplifies NVSHMEM's
waiting cost: with 16
live values,
separately compiled NVSHMEM's waiting loss grows from 7\% to 20\% and
64\% at 16, 64, and 528 RC QPs per PE.

\parheading{In DeepEP, issuing finishes early.}
In DeepEP V1's intact low-latency kernels on eight P-IB GPUs, dispatch
finishes issuing about \SI{25}{\micro\second} into a 0.49\,ms
operation; the rest is receive-side waiting and copying. Combine also
finishes issuing early, then waits at grid synchronization\append{
(Appendix~\ref{app:kernel-details})}. Reducing issue cost alone
therefore addresses only a small part of their execution.

\parheading{Takeaway.} A kernel can pay for communication before it
sends a message. Library packaging, available overlap, and completion
scope determine the cost to useful work.

\begin{figure*}[t]
  \centering
  \includegraphics[width=\textwidth]{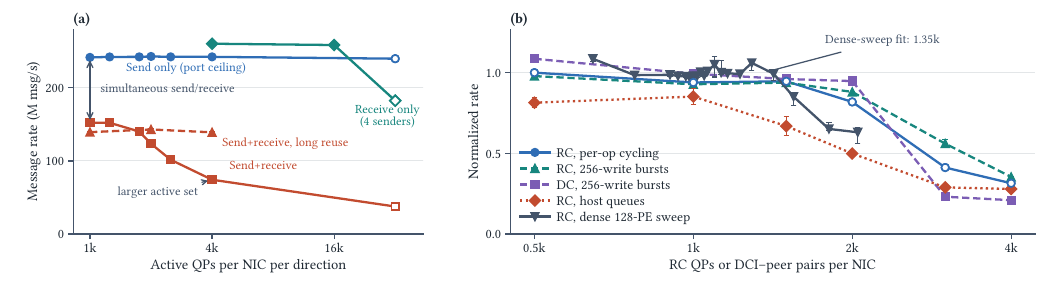}
  \caption{Connection scaling on P-H100.
    (a)~8\,B rate through one NIC against its active QPs per direction.
    Long reuse increases writes per connection visit
    from 32 to 8,192. Receive-only uses total incoming QPs and aggregate
    rate over a common duration; other curves use the median sender.
    (b)~32-PE RC/DC and dense 128-PE RC sweeps, normalized to their
    two-peer controls within each pass. The x axis counts visited QPs
    or DCI--peer pairs.
    Medians over passes with their range; hollow markers are single-pass
  points; (b) uses the mean over NICs.}
  \label{fig:icm-knee}
\end{figure*}

\subsubsection{\textbf{What does the NIC pay? Sending and serving
many connections}}
\label{sec:micro:icm}
\label{sec:micro:dc}

% \parheading{Question.}
RC connections keep per-connection QP state in the NIC's Interconnect Context
Memory (ICM), backed by host memory and cached on the NIC, alongside
completion and memory-translation state. How does message rate
change as the active connection set grows, does it matter whether the
NIC sends, receives, or both, and can connection reuse or DC reduce the cost?

\parheading{Setup.} On P-H100, GPU threads issue RDMA writes through
GPU-resident queues while a host handler rings the doorbells. We select active
QP subsets from preallocated pools at a fixed GPU grid. Eight PEs on eight
nodes exercise three roles: one sender, a receive-only target, or every
NIC sending and receiving. Unless varied, each warp posts 32 writes per
connection visit. We measure throughput including the final completion
drain, vary payload and connection reuse, and corroborate the results
with 32--128-PE all-to-all sweeps and host verbs.

\parheading{Sending and receiving together drives the loss.}
A send-only NIC sustains its rate through much larger active sets,
staying near 242\,M\,msg/s as the set of active QPs grows to 32,768.
When every NIC also receives, rate falls from 152 to 74\,M\,msg/s
over the same range (Figure~\ref{fig:icm-knee}a). Simultaneous traffic
pays an initial penalty and loses a further 51\% as the active set grows.
Receive-only traffic sits between the two, declining only at about ten
times larger active sets, and sooner with more senders.
A connection count that is inexpensive for one traffic role or workload can be
costly for another.

\parheading{Connection reuse and payload change the outcome.}
Holding each connection for 8,192 writes per warp visit flattens the
simultaneous-traffic curve to 139--143\,M\,msg/s across 1,024--4,096
QPs (Figure~\ref{fig:icm-knee}a). Reuse removes much of the growing-set
penalty, although the cost of simultaneous traffic remains.
Conversely, changing connection on
every write also hurts send-only traffic at large sets. Larger
payloads hide the message-rate
loss: at 4,096 QPs, simultaneous traffic retains only 55\% of
send-only goodput with 128\,B writes but 98\% from 256\,B
onward\append{ (Appendix~\ref{app:knee-controls})}.

\parheading{All-to-all loses 59\% by 3,000 active connections.}
The dense 128-PE RC sweep places the fitted onset near 1,350
active connections per NIC. In the 32-PE sweep,
cycling 31 peers loses 59\% of the two-peer rate by roughly 3,000
connections (Figure~\ref{fig:icm-knee}b).

\parheading{DC does not eliminate the loss.}
With 256 writes per destination, DC retains 95\% of its
two-peer rate around 1,984 DCI--peer pairs, but 23\% at 2,976; RC with the same
bursts retains 56\% there (Figure~\ref{fig:icm-knee}b). Varying DCTs
per PE from 1 to 8 does not remove this decline.

DC itself is cheap per operation, \SI{0.45}{\micro\second} above RC through
completion in mini-gda, but switching destination on every WQE costs
NVSHMEM's DC path $60\times$ in message rate on P-H100, most of which
short per-destination bursts recover. The decline itself persists with
host-resident queues, other queue depths, and host verbs without a
GPU, and packet counters show no retransmission-driven amplification
but cannot separate context-cache misses, packet-processing
contention, and backpressure\append{ (Appendix~\ref{app:knee-controls})}.

\parheading{Takeaway.} Traffic role, connection reuse, and payload
determine the cost of active connections. With one PE per NIC, cycling
16 QPs across 127 peers already exceeds the all-to-all onset, even
though a send-only NIC can sustain much larger sets. Queue count
alone is therefore insufficient to budget NIC capacity.

\section{Related Work}
\label{sec:Related-Work}

\parheading{GPU Communication Studies.}
Several recent works explore GPU communication stacks.
Demystifying NCCL \cite{demystifying-nccl} analyzes NCCL's protocols
and algorithms, and Demystifying NVSHMEM \cite{demistifying-nvshmem}
analyzes symmetric memory, transport selection, device-side RMA,
collective algorithms, and DeepEP integration. The NCCL GIN paper
\cite{gpuinitiatednetworkingnccl} describes its device API and its GPU
and proxy backends, focusing on the library rather than the mechanism
beneath it. The Landscape of GPU-Centric Communication
\cite{Landscape_Of_GPU_Comm} surveys the field and establishes the
initiation-vs.-submission taxonomy we adopt.
Shen et al.\ \cite{sol-collectives} take the same mechanism-first
approach inside an NVLink domain, deriving a speed-of-light bound
from fence and remote-store latency and showing that synchronization,
rather than data movement,
dominates small-message collectives.
Our work complements these studies by analyzing communication at the
mechanism level through minimal GPU-submitted and proxy transports that vary WQE
construction, ordering, queue sharing, and completion independently.

\parheading{GPU-Initiated Communication in HPC.}
Agostini et al.\ \cite{GPUDirect-Async} explore GPUDirect Async, where
GPUs trigger InfiniBand communication that the CPU prepares, and
Hamidouche and LeBeane \cite{rocshmem-ppopp} build a GPU-initiated
OpenSHMEM that handles the GPU--NIC consistency problem of long-running
kernels. Ismayilov et al.\ \cite{multi-gpu-iterative-solvers} propose
a CPU-free execution model for iterative solvers, and Trotter et
al.\ \cite{trotter2025cpu} compare CPU- and GPU-initiated communication
for conjugate gradient solvers on GPU clusters. Baydamirli et
al.\ \cite{autonomous-multi-gpu-compiler} extend this model with
compiler support that automatically transforms multi-GPU code to run
without CPU orchestration. These studies
evaluate communication strategies within applications; we measure the
mechanisms those strategies rely on.

\parheading{RDMA Performance Engineering.}
Kalia et al.\ \cite{kalia-atc16} established several mechanisms as host-side
optimizations, including doorbell batching, compact WQEs, payload inlining,
and unsignaled completions, some of which have been adapted into GPU-side
RDMA that we analyze in this work. Kong et al.\ \cite{rdma-microarchitecture}
characterize NIC-side resources including QP context, address translation, and
WQE cache. They find that large QP counts can pressure NIC-side caches and
degrade scaling, which motivates our active-connection experiment
(\S\ref{sec:micro:icm}) for GPU-driven traffic. FaRM shares connections
among threads to limit this per-connection state \cite{farm}.
ScaleRPC finds that
inbound and outbound WRITEs scale differently and groups connections
for locality~\cite{scalerpc}, and Collie reports bidirectional RDMA
degradation from internal packet-processing contention~\cite{collie};
our experiment separates traffic roles on ConnectX-7 and adds connection
reuse and payload as variables.

\parheading{Expert-Parallelism Libraries.}
DeepEP \cite{deepep2025} introduced IBGDA-based MoE communication and
remains the reference implementation, and pplx-kernels
\cite{pplx-kernels,pplx-blog} implements dispatch and combine over
NVSHMEM. Hybrid-EP \cite{hybrid-ep}, NIXL \cite{nixl}, and NCCL EP
\cite{nccl-ep} bring these designs into NVIDIA's Megatron, Dynamo, and
NCCL ecosystems, respectively. In contrast, fabric-lib
\cite{pplx-garden-fabric-lib}, UCCL-EP \cite{uccl-ep}, and NCCLX
\cite{ncclx} route GPU requests through host threads for portability
across GPU and NIC architectures.
These libraries are the consumers of
the mechanisms this paper dissects, and our measurements explain
several of the differences in their published performance.

\parheading{Beyond NVIDIA GPUs and ConnectX NICs.}
GPU-initiated communication is not exclusive to NVIDIA GPUs and
ConnectX NICs. AMD's rocSHMEM has GPU-submitting (GPUDirect Async)
backends for ConnectX-7, Broadcom Thor 2, and Pensando Pollara NICs
\cite{rocshmem-providers}. These backends, like IBGDA and GDAKI,
submit from GPU code rather than only letting the NIC reach GPU
buffers by DMA. The ROCm port of DeepEP drives ConnectX NICs
from AMD GPUs through the same \code{mlx5} doorbell path
\cite{rocm-deepep}, and the MORI interface provides modular RDMA
primitives for MI-series GPUs \cite{mori}. Intel SHMEM provides a
device-initiated OpenSHMEM API for Intel GPUs \cite{intel-ishmem}.

On the NIC side, AWS EFA DP Direct posts work requests from CUDA,
updating a 32-bit producer-index doorbell, rather than \code{mlx5}'s
control-segment store, between system-scope fences
\cite{efa-dp-direct,efa-device-source}, and HPE
Slingshot lets a GPU trigger NIC commands that the host prepared
\cite{slingshot-gpu-nic}. The mechanisms we measure (WQE construction,
ordered doorbells, completion scope, and NIC connection state) apply to
all of these paths, though their costs differ.

\section{Conclusion}
\label{sec:Conclusion}

This paper dissected GPU-initiated communication at the GPU--NIC
boundary. Using two minimal transports, mini-gda and mini-proxy, we
separated the costs of the hardware mechanism from those of the
libraries built on it, and measured both on four NVIDIA platforms. Our
experiments connect queue management, ordering, and completion choices
to their costs: the mechanism itself is cheap, and the choices a
library makes around it determine latency, message rate, and the
resources the kernel and the NIC must provide. We close with guidelines
for library designers and for those who benchmark them.

\begin{itemize}
  \item \textbf{Price the completion needed.} Choose completion scope
    independently of the queues needed for throughput. A PE-wide quiet
    walks every configured queue, while a signal or per-QP completion
    stays flat, so compare paths at the same queue count.
  \item \textbf{Report the processor state.} GPU submission depends on
    SM clock rate, while proxies depend on host operating state;
    report both for fair benchmarks and reproducibility.
  \item \textbf{Provision queues for isolation and capacity.} Reserve
    queues for latency-sensitive traffic in either design, and use
    batching and parallel submitters to raise capacity. A separate
    proxy ring isolates traffic even when it shares a worker.
  \item \textbf{Compile for the final caller.} Compile and measure
    communication in its final caller, where even dormant code can
    reduce useful throughput. Inspect register and stack usage for the
    actual kernel and block size, since neither inlining nor separate
    compilation is a universal default.
  \item \textbf{Budget connections by traffic, not by count.} Budget
    active connections by traffic role, reuse, and payload, which
    determine their cost to the NIC. DC reduces the number of
    persistent queues but does not remove this cost.
\end{itemize}

Understanding and
using GPU-initiated communication efficiently demands expert
knowledge and extensive testing, even inside vendor libraries. By
dissecting the GPU--NIC boundary and measuring each mechanism in
isolation, this paper aims to give users and developers of communication
libraries an explainable reference point and a performance oracle for
the transport.
We release mini-gda, mini-proxy, and the benchmark suite as open source at \GitHubRepoURL.

% ================================================================

\section*{Acknowledgments}
Authors from Koç University have received funding from
the European Research Council (ERC) under the European
Union's Horizon 2020 research and innovation programme
(grant agreement No 949587).
We acknowledge the EuroHPC Joint Undertaking for awarding access to
the MareNostrum5 supercomputer in Spain, and TÜBİTAK ULAKBİM, High
Performance and Grid Computing Center (TRUBA resources), where the
experiments reported in this paper were partially performed.

% References
\bibliographystyle{\paperbibliographystyle}
\bibliography{references}

% Appendix
\clearpage
\ifpaperieee
\appendices
\else
\appendix
\fi
\raggedbottom

\section{Methodology Details}
\label{app:methodology}
As a timer check, adding a \SI{50}{\micro\second} responder delay
shifts round trips by 49.6--\SI{50.3}{\micro\second} on both P-IB and
P-GB200. The SM clock measured 1,962--1,980\,MHz during P-IB latency
runs and 2,062\,MHz on P-GB200. Runs discard 100--1,000 warm-up
operations. Section~\ref{sec:micro:doorbell-issue-cost} and the idle,
handoff, message-rate, and payload measurements of
Section~\ref{sec:micro:proxydesign} use one P-IB pair; the loaded-RTT
curves, including their idle controls, come from a second pair.

\section{Single-Operation Controls}
\subsection{SM-clock sensitivity}
\label{app:clockfit}
The P-RoCE sweep uses NVSHMEM 3.7.2 with two RC QPs per peer and
NCCL 2.31.2 GDAKI with one context. Three shuffled passes measure
5,000 operations per cell at six SM clocks (502--1,845\,MHz);
payload and NIC packet-count checks pass.

\begin{table}[H]
  \centering
  \caption{Selected P-RoCE fits, $T(f)=C/f+B$, using six clock medians,
    with $T$ in \si{\micro\second} and $f$ in MHz. $C$ is an effective
  clock-sensitive coefficient.}
  \label{tab:clockfit}
  \begin{tabularx}{\columnwidth}{@{}Y S[table-format=5.0]
    S[table-format=2.2] S[table-format=1.4]@{}}
    \toprule
    \textbf{Sequence} & {\textbf{$\boldsymbol{C}$ (cycles)}} &
    {\textbf{$\boldsymbol{B}$ (\si{\micro\second})}} &
    {\textbf{$\boldsymbol{R^2}$}} \\
    \midrule
    NVSHMEM public issue & 10313 & 1.44 & 0.9970 \\
    NVSHMEM public put+completion & 16987 & 12.39 & 0.9963 \\
    GDAKI inline issue & 2929 & 0.27 & 0.9977 \\
    GDAKI inline put+flush & 5683 & 10.30 & 0.9432 \\
    IBRC enqueue & 1891 & 0.09 & 0.9986 \\
    IBRC put+completion & 6045 & 8.91 & 0.9514 \\
    \bottomrule
  \end{tabularx}
\end{table}

\subsection{Completion scope}
\label{app:quiet-tax}
The P-RoCE scope experiment uses three shuffled passes of 5,000
operations after 1,000 warm-up operations at a measured 1,965\,MHz.
Both used-QP routines complete the QP selected by the same internal
put.

\begin{table}[H]
  \centering
  \caption{P-RoCE completion scope, NVSHMEM 3.7.2: 8\,B latency.
    All-QP, DeepEP \emph{port}, and NVSHMEM \emph{stock}
    completion use the same internal put. DeepEP and stock complete
  only its selected QP. RTT is a signal-based control.}
  \label{tab:proce-scope}
  \begin{tabular}{@{}r S[table-format=2.2] S[table-format=2.2]
    S[table-format=2.2] S[table-format=2.2]@{}}
    \toprule
    & \multicolumn{3}{c}{\textbf{Put+completion (\si{\micro\second})}} & \\
    \cmidrule(lr){2-4}
    \textbf{QPs/peer} & {\textbf{All QPs}} & {\textbf{Used, port}} &
    {\textbf{Used, stock}} & {\textbf{RTT (\si{\micro\second})}} \\
    \midrule
    1  & 18.02 & 16.93 & 17.25 & 31.17 \\
    2  & 18.78 & 16.90 & 17.09 & 31.78 \\
    4  & 19.78 & 15.14 & 17.98 & 31.94 \\
    8  & 20.96 & 15.87 & 17.60 & 31.58 \\
    16 & 24.77 & 17.09 & 17.92 & 30.98 \\
    32 & 34.85 & 16.93 & 15.49 & 29.98 \\
    \bottomrule
  \end{tabular}
\end{table}

\begin{table}[H]
  \centering
  \caption{P-GB200 NVSHMEM 3.7.2 put+completion.
    DCI requested $=0$ selects automatic sizing. One GPU and NIC per
  node, all traffic through the NIC. SM clock at 2,062\,MHz.}
  \label{tab:gb200-scope}
  \begin{tabular}{@{}r r r S[table-format=2.2] S[table-format=2.2]@{}}
    \toprule
    & & & \multicolumn{2}{c@{}}{\textbf{Put+completion (\si{\micro\second})}} \\
    \cmidrule(l){4-5}
    \textbf{RC/peer} & \textbf{DCI requested} & \textbf{DCI actual} &
    {\textbf{Public}} & {\textbf{Internal}} \\
    \midrule
    2  & 1 & 1   & 13.86 & 12.19 \\
    16 & 1 & 1   & 22.98 & 21.28 \\
    2  & 0 & 153 & 94.88 & 92.86 \\
    \bottomrule
  \end{tabular}
\end{table}

\section{Proxy Controls}

\subsection{Operating state}
\label{app:proxy-states}
\label{app:pairs}
Cold workers follow 45\,s idle (2.5\,GHz); warm workers follow 20\,s
sustained load (4.0\,GHz). CPU-only preconditioning reproduces the
warm results while GPU-submitted controls change little.

\begin{table}[H]
  \centering
  \caption{P-IB proxy latency, measured cold and warm in one
  session on one pair.}
  \label{tab:proxy-states}
  \setlength{\tabcolsep}{2.5pt}
  \begin{tabular}{@{}l S[table-format=1.2] S[table-format=1.2]
      S[table-format=2.2] S[table-format=2.2] S[table-format=2.2]
    S[table-format=2.2]@{}}
    \toprule
    & \multicolumn{2}{c}{\textbf{Put+c. (\si{\micro\second})}} &
    \multicolumn{2}{c}{\textbf{RTT (\si{\micro\second})}}
    & \multicolumn{2}{c@{}}{\textbf{RTT p99 (\si{\micro\second})}} \\
    \cmidrule(lr){2-3} \cmidrule(lr){4-5} \cmidrule(l){6-7}
    \textbf{Path} & {\textbf{Cold}} & {\textbf{Warm}} &
    {\textbf{Cold}} & {\textbf{Warm}} & {\textbf{Cold}} & {\textbf{Warm}} \\
    \midrule
    IBRC & 6.40 & 6.30 & 11.07 & 10.69 & 12.80 & 11.46 \\
    GIN Proxy & 8.19 & 7.57 & 16.32 & 14.53 & 16.86 & 15.07 \\
    mini-proxy, baseline & 7.07 & 6.91 & 14.59 & 13.44 & 16.93 & 13.86 \\
    mini-proxy, tuned & 4.10 & 3.62 & 5.89 & 5.22 & 8.86 & 5.57 \\
    \bottomrule
  \end{tabular}
\end{table}

\begin{table}[H]
  \centering
  \caption{Selected P-IB proxy capacity.}
  \label{tab:proxy-capacity-details}
  \begin{tabular}{@{}l r S[table-format=2.1] S[table-format=2.1]@{}}
    \toprule
    & & \multicolumn{2}{c@{}}{\textbf{M\,msg/s}} \\
    \cmidrule(l){3-4}
    \textbf{Path} & \textbf{Workers} & {\textbf{Cold}} & {\textbf{Warm}} \\
    \midrule
    mini-proxy, $B=1$ & 1 & 2.6 & 6.6 \\
    mini-proxy, $B=16$ & 1 & 14.4 & 28.6 \\
    UCCL-EP & 8 & 34.0 & 60.8 \\
    MSCCL++, patched & 8 & 7.5 & 10.2 \\
    GIN Proxy, 4 contexts & 4 & 3.3 & 7.0 \\
    fabric-lib & 1 & 8.6 & 12.6 \\
    \bottomrule
  \end{tabular}
\end{table}

\parheading{MSCCL++ Patch.}
\label{app:msccl-artifact}
Passing memory handles by const reference removes shared-pointer
refcount contention, raising P-IB capacity from 2.5 to 7.5\,M\,msg/s
over one to eight services. The patched build retains one signaled
post per message and is used throughout the paper.

\subsection{Queue isolation}
\label{app:proxy-details}

\begin{table}[H]
  \centering
  \caption{Loaded 8\,B RTT p50, 64 background CTAs.
    GDAKI uses eight threads per CTA and a fixed 65-context pool:
    the shared probe uses one
    background CTA's context. mini-proxy uses T4/R32/B16; its shared/private
    initiator rates are 47.5/44.9\,M\,msg/s on P-IB and 61.8/62.9 on P-GB200.
    UCCL-EP uses eight workers. Paced rows compare reserved queues at
  approximately 30\,M\,msg/s.}
  \label{tab:proxy-controls}
  \begin{tabularx}{\columnwidth}{@{}Y S[table-format=5.1]
    S[table-format=3.1]@{}}
    \toprule
    & \multicolumn{2}{c@{}}{\textbf{RTT (\si{\micro\second})}} \\
    \cmidrule(l){2-3}
    \textbf{Path} & {\textbf{Shared}} & {\textbf{Reserved}} \\
    \midrule
    \tabgroup{3}{Unpaced}
    GDAKI, P-IB & 27.3 & 13.9 \\
    mini-proxy, P-IB & 2190 & 115 \\
    mini-proxy, P-GB200 & 1401 & 178 \\
    UCCL-EP, P-IB & 7570 & 120 \\
    MSCCL++, P-IB & 50800 & 825 \\
    \midrule
    \tabgroup{3}{Paced to about 30\,M\,msg/s}
    GDAKI, P-IB & {---} & 12.9 \\
    mini-proxy, P-IB & {---} & 29.2 \\
    \bottomrule
  \end{tabularx}
\end{table}

\subsection{Payload size}
\label{app:payload-latency}

\begin{figure}[H]
  \centering
  \includegraphics[width=\columnwidth]{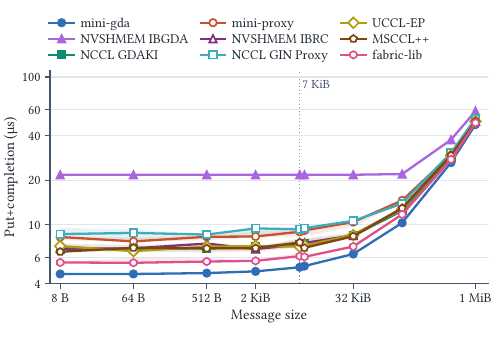}
  \caption{Single-operation put+completion against payload size on
    P-IB, for the paths of Figure~\ref{fig:crossover-proxies}.
  NVSHMEM uses all-QP quiet.}
  \label{fig:payload-latency}
\end{figure}

\begin{table}[H]
  \centering
  \caption{Selected P-GB200 pointer-write goodput, with four
    CTAs at 4\,KiB and one at 64\,KiB. mini-proxy uses T4/R32/B16,
    IBGDA 16 RC QPs, GIN Proxy four requested workers and one context
    per CTA, and GDAKI 32 threads and one context per CTA.
  Coarse sizes and changing geometry do not locate a crossover.}
  \label{tab:gb200-payload}
  \begin{tabular}{@{}l S[table-format=2.1] S[table-format=2.1]@{}}
    \toprule
    & \multicolumn{2}{c@{}}{\textbf{Goodput (GB/s)}} \\
    \cmidrule(l){2-3}
    \textbf{Path} & {\textbf{4\,KiB}} & {\textbf{64\,KiB}} \\
    \midrule
    mini-proxy & 49.1 & 49.0 \\
    NVSHMEM IBGDA & 49.3 & 49.5 \\
    GIN Proxy & 47.9 & 48.5 \\
    GDAKI & 21.3 & 48.0 \\
    \bottomrule
  \end{tabular}
\end{table}

\section{Resource and Scaling Controls}

\subsection{Queue publication}
\label{app:publication}
Lane drift also explains a warm-up sensitivity on P-GB200: lengthening
the warm-up from 100 to 1,000 iterations cuts GDAKI's aggregated rate
from 67 to 17\,M\,msg/s, while synchronizing after each put keeps it
at 82\,M\,msg/s in both cases (Table~\ref{tab:publication-controls}).

\begin{table}[H]
  \centering
  \caption{Selected 8\,B message-rate controls.
    P-IB GDAKI uses 16 contexts, one per CTA; synchronizing after
    each put raises the 32-thread case to 106--110\,M\,msg/s.
    NVSHMEM and mini-gda use 256 threads per CTA and one QP per CTA;
    mini-gda publishes cooperatively with 32 WQEs per doorbell.
    P-GB200 GDAKI uses 16 contexts of 32 threads, aggregates every
  16 puts, disables NCCL's RAS monitoring, and measures 10,000 iterations.}
  \label{tab:publication-controls}
  \begin{tabularx}{\columnwidth}{@{}Y r r@{}}
    \toprule
    \textbf{P-IB configuration} & \multicolumn{2}{r@{}}{\textbf{M\,msg/s}} \\
    \midrule
    GDAKI, 1 / 8 threads per context & \multicolumn{2}{r@{}}{12.0 / 39.5} \\
    GDAKI, 32 / 64 threads per context & \multicolumn{2}{r@{}}{34.3 / 27.2} \\
    NVSHMEM, 2 / 16 RC QPs & \multicolumn{2}{r@{}}{50.2 / 251.5} \\
    mini-gda, 16 QPs / 4,096 threads & \multicolumn{2}{r@{}}{260} \\
    \midrule
    \textbf{P-GB200 warm-up iterations} & \textbf{Original} &
    \textbf{Sync after put} \\
    \midrule
    100 & 67.10 & 82.20 \\
    1,000 & 16.61 & 82.12 \\
    \bottomrule
  \end{tabularx}
\end{table}

\subsection{Enclosing-kernel resources}
\label{app:kernel-details}
Kernel times are medians of three paired passes of 15 launches. The
inlined build sets \code{NVSHMEM_ENABLE_ALL_DEVICE_INLINING=ON}.

\begin{table}[H]
  \centering
  \caption{P-IB compute caller with eight live values
    (\code{sm_90}, \code{-O3}). Waiting loss is relative to sending
  without waiting, with 16 individual 8\,B messages per block.}
  \label{tab:kernel-details}
  \begin{tabular}{@{}l r r S[table-format=2.1]@{}}
    \toprule
    & \multicolumn{2}{c}{\textbf{Registers/thread}} & \\
    \cmidrule(lr){2-3}
    \textbf{Path} & \textbf{Absent} & \textbf{Dormant} &
    {\textbf{Waiting loss (\%)}} \\
    \midrule
    mini-gda & 29 & 32 & 1.9 \\
    mini-proxy & 29 & 34 & 26.8 \\
    NVSHMEM, separate & 34 & 44 & 22.3 \\
    NVSHMEM, inlined & 34 & 98 & 18.8 \\
    GDAKI & 29 & 95 & 5.3 \\
    \bottomrule
  \end{tabular}
\end{table}

With dormant NVSHMEM code and the earlier instrumented builds, a
compute caller with 160 live values retains 15\% of its throughput
when NVSHMEM is compiled separately and 100\% when it is inlined; at
224 live values, it retains 8\% and 25\%.

\parheading{DeepEP Configuration.}
DeepEP V1 (NVSHMEM 3.4.5) uses two four-GPU P-IB nodes, 128 tokens
per rank, hidden size 7,168, 288 experts, top-8 routing, and all-RDMA
traffic. Across five runs, combine finishes issuing after
\SI{24}{\micro\second}, and its median warp spends 0.93\,ms at grid
synchronization before reduction.

\subsection{Active connections}
\label{app:knee-controls}
The dense 128-PE sweep fits a plateau
joined to a power-law decline by least squares in log space.

\begin{table}[H]
  \centering
  \caption{P-H100 active-connection controls. GPU runs use CPU-forwarded
    doorbells and normalize to each configuration's two-peer control.
  Host verbs normalize to their one-sender control.}
  \label{tab:knee-controls}
  \begin{tabularx}{\columnwidth}{@{}>{\raggedright\arraybackslash}p{0.35\columnwidth}
    Y@{}}
    \toprule
    \textbf{Control} & \textbf{Observation} \\
    \midrule
    RC, host / GPU queues &
    0.50 / 0.82 at 1,984 connections; 0.29 / 0.41 at 2,976 \\
    Queue depth & Rate changes within $\pm2\%$ \\
    DC, 1--8 DCTs per PE & 0.95 at 1,984 DCI--peer pairs;
    0.23--0.24 at 2,976 \\
    Host verbs, send + receive & 0.60--0.75 at 1,024 QPs/direction;
    $\leq0.10$ from 2,048 \\
    Host verbs, one sender & Flat through 7,168 QPs \\
    Packets, send + receive &
    1.01--1.09 per message, $\leq0.1$ ACK per received message
    over 1,024--7,168 QPs \\
    \bottomrule
  \end{tabularx}
\end{table}

\begin{table}[H]
  \centering
  \caption{Selected P-H100 payload controls: 4,096 active QPs per NIC,
    four peers, 32 writes per connection visit. Goodput is the
    median sender rate per pass, then median [min, max] over passes
  (three at 8\,B, two otherwise). Ratio divides the two medians.}
  \label{tab:knee-payload}
  \begin{tabular}{@{}r r r S[table-format=1.2]@{}}
    \toprule
    & \multicolumn{2}{c}{\textbf{Goodput (GB/s)}} & \\
    \cmidrule(lr){2-3}
    \textbf{Size (B)} & \textbf{Send only} & \textbf{Send + receive}
    & {\textbf{Ratio}} \\
    \midrule
    8 & 1.94 [1.94, 1.94] & 0.59 [0.58, 0.61] & 0.31 \\
    128 & 18.12 [18.12, 18.13] & 9.90 [9.71, 10.09] & 0.55 \\
    256 & 21.00 [21.00, 21.00] & 20.65 [20.64, 20.66] & 0.98 \\
    4,096 & 24.68 [24.68, 24.68] & 24.35 [24.35, 24.35] & 0.99 \\
    \bottomrule
  \end{tabular}
\end{table}

\end{document}